\documentclass[a4paper,fleqn,usenatbib]{mnras}

\usepackage{graphicx}
\usepackage[percent]{overpic}
\usepackage{lipsum}
\usepackage{multirow}
\usepackage{color}
\usepackage{rotating}
\usepackage{mwe,tikz}
\usepackage[percent]{overpic}
\usepackage{mathtools}
\usepackage{physics}
\usepackage{amsmath}
\usepackage[utf8x]{inputenc}
\usepackage{hyperref}
\usepackage{longtable}
\usepackage{wasysym}
\usepackage[normalem]{ulem}
\hypersetup{
    colorlinks=true,
    linkcolor=blue,
    filecolor=magenta,      
    urlcolor=blue,
}
\usepackage{soul}
\usepackage{pdflscape}
\usepackage{longtable}
\usepackage{booktabs,tabularx}

\def\ltsima{$\; \buildrel < \over \sim \;$}
\def\simlt{\lower.5ex\hbox{\ltsima}}
\def\gtsima{$\; \buildrel > \over \sim \;$}
\def\simgt{\lower.5ex\hbox{\gtsima}}
\def\kms{{\rm\,km\,s^{-1}}}
\def\masyr{{\rm\,mas\,yr^{-1}}}

\def\kpc{{\rm\,kpc}}

\def\Vcirc{{\rm\,V_{circ,\odot}}}
\def\Rsun{{\rm\,R_{\odot}}}
\def\V3Dsun{{\bmath{\rm\,V_{\odot}}}}
\def\Vpsun{{\bmath{\rm\,V_{pec,\odot}}}}

\def\VRsun{{\rm\,V_{R,\odot}}}
\def\Vphisun{{\rm\,V_{\phi,\odot}}}
\def\Vzsun{{\rm\,V_{z,\odot}}}

\def\VDDLSR{{\bmath{\rm\,V_{LSR,\odot}}}}
\def\Vcirc{{\rm\,V_{circ,\odot}}}

\def\VRsunA{{\rm \, 8.88^{+1.20}_{-1.22}}}
\def\VphisunA{{\rm\,241.91^{+1.61}_{-1.73}}}
\def\VzsunA{{\rm\, 3.08^{+1.06}_{-1.10}}}

\def\vdveci{{\bmath{v}^{i}_{\rm d}}}
\def\vdvec{{\bmath{v}_{\rm d}}}
\def\vobsmui{{\bmath{v}^{i}_{\rm obs, \mu}}}
\def\vobsmu{{\bmath{v}_{\rm obs, \mu}}}
\def\vperpi{{\bmath{v}^{i}_{\rm reflex, \bot}}}
\def\vperp{{\bmath{v}_{\rm reflex, \bot}}}

\def\nstreams{{\rm\,17}}

\def\Gaia{{\it Gaia\,}}
\defcitealias{Malhan2017}{MI17}


\def\Gyr{{\rm\,Gyr}}

\def\ltsima{$\; \buildrel < \over \sim \;$}
\def\gtsima{$\; \buildrel > \over \sim \;$}

\title[Measuring the Sun's motion with streams using Gaia EDR3]{Measuring the Sun's velocity using Gaia EDR3 observations of Stellar Streams}

\author[Malhan et al.]{
Khyati Malhan$^{1}$\thanks{E-mail: khyati.malhan@fysik.su.se},
Rodrigo A. Ibata$^{2}$,
Nicolas F. Martin$^{2,3}$
\\
$^{1}$The Oskar Klein Centre, Department of Physics, Stockholm University, AlbaNova, SE-10691 Stockholm, Sweden\\
$^{2}$Universit\'e de Strasbourg, CNRS, Observatoire Astronomique de Strasbourg, UMR 7550, F-67000 Strasbourg, France\\
$^{3}$Max-Planck-Institut f\"ur Astronomie, K\"onigstuhl 17, D-69117, Heidelberg, Germany\\
}

\date{Accepted XXX. Received YYY; in original form ZZZ}
\begin{document}
\label{firstpage}
\pagerange{\pageref{firstpage}--\pageref{lastpage}}
\maketitle

\begin{abstract}
We measure the Sun's velocity with respect to the Galactic halo using \Gaia Early Data Release 3 (EDR3) observations of stellar streams. Our method relies on the fact that, in low-mass streams, the proper motion of stars should be directed along the stream structure in a non-rotating rest frame of the Galaxy, but the observed deviation arises due to the Sun's own reflex motion. This principle allows us to implement a simple geometrical procedure, which we use to analyse $\nstreams$ streams over a $\sim 3$--$30\kpc$ range. Our constraint on the Sun's motion is independent of any Galactic potential model, and it is also uncorrelated with the Sun's galactocentric distance. We infer the Sun's velocity as $\VRsun=\VRsunA\kms$ (radially towards the Galactic centre), $\Vphisun=\VphisunA\kms$ (in the direction of Galactic rotation) and $\Vzsun=\VzsunA\kms$ (vertically upwards), in global agreement with past measurements through other techniques; although we do note a small but significant difference in the $\Vzsun$ component. Some of these parameters show significant correlation and we provide our MCMC output so it can be used by the reader as an input to future works. The comparison between our Sun's velocity inference and previous results, using other reference frames, indicates that the inner Galaxy is not moving with respect to the inertial frame defined by the halo streams.
\end{abstract}

%%%%%%%%%%%%%%%%%%%%%%%%%%%%%%%%%%%%%%%%%%%%?KT2%%%%%%%%%%%%%%%%%%%%%%
%%%%%%%%%%%%%%%%% BODY OF PAPER %%%%%%%%%%%%%%%%%%%%%%%%%%%%%%%%%%
%%%%%%%%%%%%%%%%%%%%%%%%%%%%%%%%%%%%%%%%%%%%%%%%%%%%%%%%%%%%%%%%%%
\section{Introduction}\label{sec:Introduction}

The measurement of the Sun's velocity with respect to our Galaxy, $\V3Dsun$, is very fundamental to the field of astronomy. Its precise value is important, for instance, to transform any observed Heliocentric velocity into the rest frame of the Milky Way. This is necessary for scientific interpretation when studying Galactic dynamics and when correcting the motion of extragalactic systems. Moreover, the rotation component of the Sun's motion serves as a useful constraint for the mass models of the Milky Way (e.g., \citealt{DehnenBinney1998, Bovy2015}). The knowledge of the Sun's velocity is also crucial for the direct detection experiments that are dedicated to find `annual-modulation' signals from dark matter (cf. \citealt{Freese2013}). Given the fundamental significance of this astronomical parameter, it is surprising that we do not possess any independent measurement for which all the three components of the Sun's galactic velocity ($\V3Dsun \equiv \VRsun,\Vphisun,\Vzsun)$\footnote{Galactic coordinates are aligned such that $\VRsun$ is in the direction to the Galactic centre, $\Vphisun$ is the total rotational velocity in the direction of the local disk rotation, and $\Vzsun$ is in the direction perpendicular to the Galactic disk.} were simultaneously determined.

Generally, the three-dimensional vector $\V3Dsun$ is required to be ``constructed'' by combining the measurements from different studies, where each study would have measured different component(s) of $\V3Dsun$. In practice, this means combining two separate measurements. (1) The Sun's rotational velocity around the Galaxy, which is measured either with respect to the Galactic centre (e.g., inferred from ``short-period'' stars orbiting the black hole Sagittarius $\rm{A^*}$, \citealt{Reid2004, Ghez2008}) or in the Galactic disk (inferred from stars or star-forming regions, \citealt{Reid2009, Bovy2009, McMillan2010, Bovy2012, Eilers2019}). These measurements either provide the $\Vphisun$ component, or the circular component of $\Vphisun$ (i.e., $\Vcirc$). (2) The Sun's velocity with respect to the solar neighbourhood (that defines the Sun's ``peculiar velocity'' with respect to the Local Standard of Rest, LSR, \citealt{DehnenBinney1998_Sun, Schonrich2010, Bobylev2017, Kawata2019}). This provides the $\VRsun$ and $\Vzsun$ components. However, by combining different measurements from different studies in this way, one implicitly makes a set of broad assumptions. The first assumption that one makes is that these different components of $\V3Dsun$, arising from different studies, are compatible with each other, and therefore can be added vectorially. Second, one also assumes that the Milky Way disk is axisymmetric and in a  steady-state, and thus, that the Sun's total Galactic velocity can be expressed as a simple addition of the circular motion of the LSR and the Sun's peculiar motion:  $\V3Dsun=\Vcirc+\Vpsun$. But we know that the disk is not in a steady state, and in fact, is dynamically perturbed; the testimony of this are the stars in the solar neighbourhood that are oscillating in radial and vertical directions \citep{Antoja2018}. This effect can give rise to non-circular streaming motions of stars at the Sun's location. Therefore, if the solar neighbourhood (that defines the LSR) itself has a motion with respect to $\Vcirc$ (the true circular velocity at the Sun's location), then the actual decomposition would be $\V3Dsun=\Vcirc+\VDDLSR+\Vpsun$, but we have no straightforward means to determine $\VDDLSR$ (although see \citealt{Drimmel_2018}). To avoid such possible complications, it is desirable to measure the Sun's velocity using an approach that can simultaneously constrain all three components of $\V3Dsun$\footnote{To our knowledge, \cite{Drimmel_2018} is the only study that provides all the components of $\V3Dsun$. However, they achieved this by combining two different measurements: the Sun's Galactic distance $\Rsun$ (from \citealt{Gravity_Collaboration2018}) and the proper motion of Sagittarius $A^{*}$ (from \citealt{Reid2004}).}. 

An alternative way to determine the Sun's velocity is to make this measurement with respect to those tracers that orbit the Galactic halo. This scenario is less explored, but it holds the potential to measure the full vector $\V3Dsun$. Note that the Sun's motion measured with respect to the Galactic halo, in principle, could be different from that measured either with respect to the Galactic centre or the Galactic disk. This could happen if, for instance, the Milky Way halo has a bulk motion or rotation with respect to the Galactic centre (as may be happening due to the ongoing accretion of the Magellanic Clouds; e.g.,  \citealt{Petersen2020}). However, a similarity between these different measurements of the Sun's velocity would indicate that the dynamical centre of the halo has no relative motion with respect to the dynamical centre of the disk. The prospect of exploring these interesting astrophysical scenarios serves as an additional motivation for us to attempt measuring the Sun's motion relative to the Galactic halo. 

In our study, we measure the Sun's velocity with respect to the Galactic halo using {\it stellar streams}. Stellar streams are the remains of satellites that have been disrupted by the tidal forces of the host galaxy. In particular, a stream that is produced by a low-mass satellite (e.g., a star cluster) follows closely the orbit of its parent satellite. Therefore, a low-mass stream closely delineates orbit in the Galactic potential (\citealt{Dehnen2004, Varghese2011, Bowden2015}). It is this orbital behaviour of streams that we exploit to measure the Sun's velocity $\V3Dsun$. As described below, the main advantage of our technique is that the resulting measurement of $\V3Dsun$ is independent of any Galactic potential model, and it is also uncorrelated with the value of the Sun's Galactic distance $\Rsun$; which is often not the case in previous studies. 

The paper is arranged as follows. In Section~\ref{sec:Method}, we explain our method for measuring the Sun's velocity using stellar streams. Section~\ref{sec:data} details the observational data used, and Section~\ref{sec:analysis} describes the analysis performed to determine the Sun's velocity. Finally, we present our conclusions and discussion in Section~\ref{sec:conclusion_and_discussion}.

\section{Method}\label{sec:Method}

\begin{figure}
\begin{center}
\includegraphics[width=\hsize]{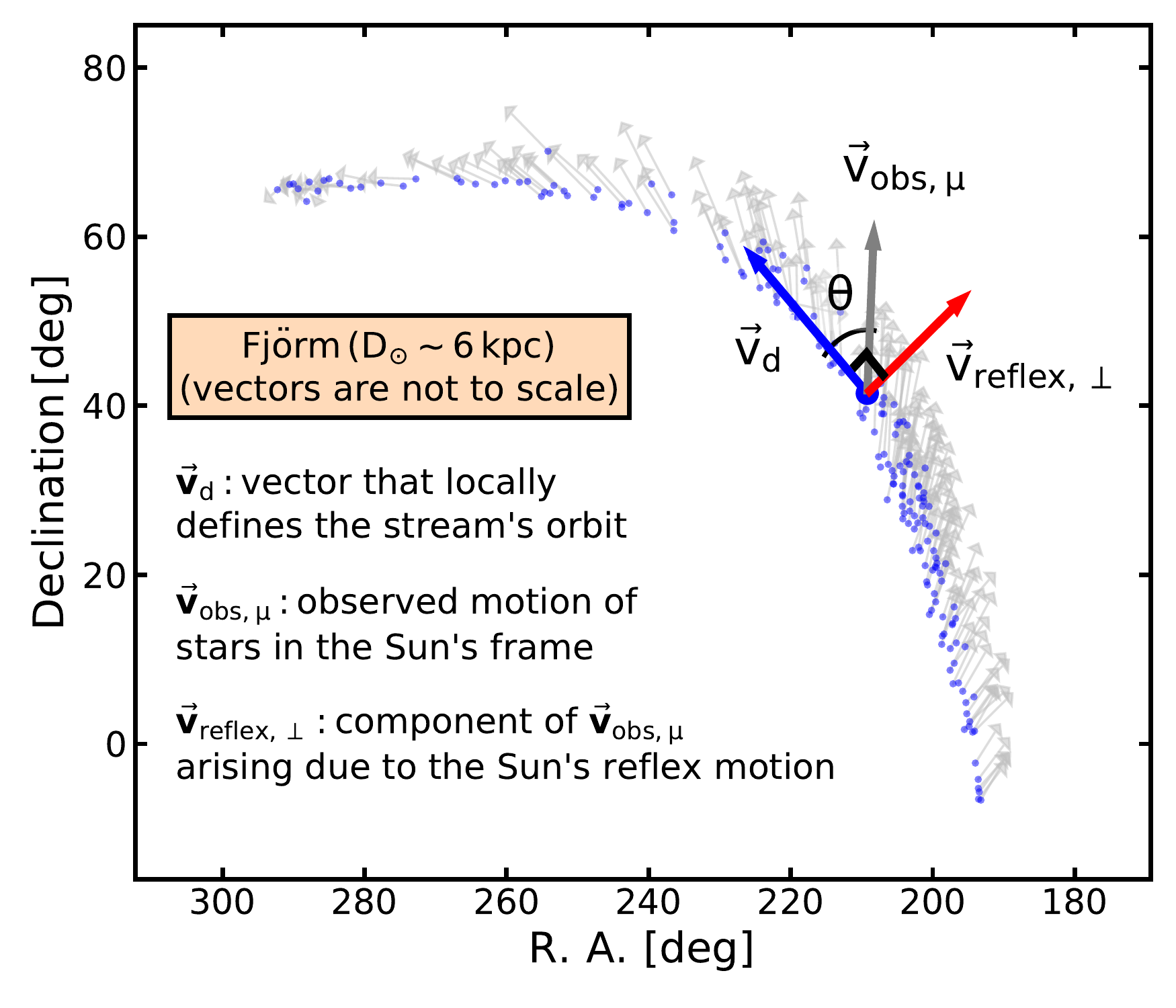}
\end{center}
\vspace{-0.5cm}
\caption{Schematic diagram based on the measurements of the ``Fj\"orm'' stream \citep{Ibata_norse_2019}. The blue points represent the positions of stars, and these reveal the stream's structure on this 2D celestial sky. The gray arrows represent the measured proper motions of stars. $\vdvec$ is a vector that is tangent to the stream's structure, and it is defined locally at every stellar point using the positions. However, it is shown here only for a specific star for the purpose of demonstration. All the $\vdvec$'s, together, trace the orbit of the stream on this 2D sky. Vector $\vobsmu$ is constructed for every star using the proper motions. Vector $\vperp$ is a component of $\vobsmu$, and it is perpendicular to $\vdvec$. Our hypothesis is that $\vperp$ primarily arises due to the Sun's own reflex motion, and thus, it holds the information about the Sun's Galactic velocity.}
\label{fig:Fig_schematic}
\end{figure}

To measure the Sun's Galactic velocity, we exploit the orbital property of stellar streams. Streams closely delineate orbits in the Galactic potential. This implies that if a stream is observed in a non-rotating rest frame of the Galaxy, then the measured velocity vectors of its stars should align along the stream structure itself (by the definition of an ``orbit''). However, in reality, all the observations are made in the Galaxy's rotating frame (i.e., the Sun's frame), and thus, the measured velocities of stream stars ``appear'' to be systematically misaligned from the stream structure. This is shown in the schematic diagram in Figure~\ref{fig:Fig_schematic} that presents the observed positions and proper motions of a Milky Way stream. The figure instantly reveals the misalignment between the stream's structure (that can be traced along the stellar positions) and the measured proper motions of stars (denoted with arrows). The primary reason for this misalignment is the Sun's own reflex motion that gives rise to the ``apparent'' motion of stars perpendicular to the stream. Thus, our working hypothesis is that the component of the observed proper motion, which is perpendicular to the stream structure, arises entirely due to the Sun's own reflex motion. Therefore, our overall strategy narrows down to first computing these perpendicular components in a sample of streams, and then using this information to determine the Sun's velocity. To this end, we adopt the following procedure.

For a given stream, we first approximate its orbit on the 2D celestial sky using the positions of stars. For this, at every stellar point, we define a local tangent vector $\vdvec$ as
\begin{equation}\label{eq:vd}
	\vdveci = \cos \delta^{i} (\alpha^{i+1}-\alpha^{i})\, \bmath{\hat{\alpha}} + (\delta^{i+1}-\delta^{i}) \, \bmath{\hat{\delta}} \, ,
\end{equation}
where  $(\alpha,\delta)$ are the R.A. and Declination of the stars, $(\bmath{\hat{\alpha}},\bmath{\hat{\delta}})$ are the corresponding unit vectors, $i$ represents a given star and $i+1$ is the next star along the stream. The series of these local tangent vectors trace out the stream's orbit on the sky.

Next, we construct the observed proper motion vector for every star  as 
\begin{equation}\label{eq:vobsmu}
\vobsmui= \mu^{*,i}_{\alpha}\, \bmath{\hat{\alpha}} + \mu^{i}_{\delta}\, \bmath{\hat{\delta}} \,,
\end{equation}
where $\mu^{*}_{\alpha} (\equiv \mu_{\alpha} \cos \delta)$ and $\mu_{\delta}$ are the measured proper motion components along R.A. and Declination, respectively.

Now, $\vobsmu$ would have aligned along $\vdvec$, had the stream been observed in the rest frame of the Galaxy. However, since the measurements are made in the moving Heliocentric frame, $\vobsmu$ and $\vdvec$ do not align (as can be seen in Fig~\ref{fig:Fig_schematic}). This occurs because the reflex motion of the Sun gives rise to the ``apparent'' motion of stars perpendicular to the stream. This component of $\vobsmu$, that is perpendicular to $\vdvec$, is defined at every stellar point as 
\begin{equation}\label{eq:perp_component}
\vperpi = (|\vobsmui| \, \sin\theta^{i}) \, \bmath{\hat{v_{r}}_{\bot}} \, ,
\end{equation}
where $|\vobsmu|$ is the magnitude of the vector $\vobsmu$, $\theta$ is the angle between $\vobsmu$ and $\vdvec$, and $\bmath{\hat{v}}_{r \bot}$ is the unit vector perpendicular to $\vdvec$. It is important to note that the quantity $\vperp$ holds the information about the relative velocity between the stream stars and the Sun. For instance, if a stream has its Galactic motion in a similar direction to that of the Sun, then the corresponding vectors $\vobsmu$ and $\vdvec$ will align close to each other, and consequentially $\vperp\sim 0$. This implies that such streams will not be very useful for constraining the Sun's Galactic velocity. On the other hand, if a stream has its orbital plane perpendicular to the Galactic plane (as is the case for the Sagittarius stream), then $|\vperp| \sim |\vobsmu|$. Such streams are very useful to measure at least the rotational component $\Vphisun$ of $\V3Dsun$ \citep{Majewski2006, Hayes2018}. In our study, we are able to constrain all three components of $\V3Dsun$ because we use multiple streams with a range of orbital inclinations and motions.

In the analysis presented in Section~\ref{sec:analysis}, we perform these vector computations for a sample of streams. Finally, a comparison is made between $\vperp$ obtained from the data (as described above) and those computed from models. The model $\vperp$ are calculated using 2D positions and distances of the stream stars for various values of the Sun's velocity ($\V3Dsun \equiv \VRsun,\Vphisun,\Vzsun$). These velocity values are sampled using a Markov Chain Monte Carlo (MCMC) algorithm. To finally gauge $\V3Dsun$, the figure of merit is adopted to be the likelihood of the data given the model. 

The overall success of this method in measuring the Sun's velocity has already been tested in \citet[hereafter \citetalias{Malhan2017}]{Malhan2017} using realistic N-body models of streams. A clear advantage of this technique is that it simply exploits the geometry of streams and does not make any assumptions about the Galactic potential of the Milky Way. Furthermore, our method does not require the radial velocity information of stars (which we generally lack for a majority of tracers).

\begin{figure*}
\begin{center}
\includegraphics[width=\hsize]{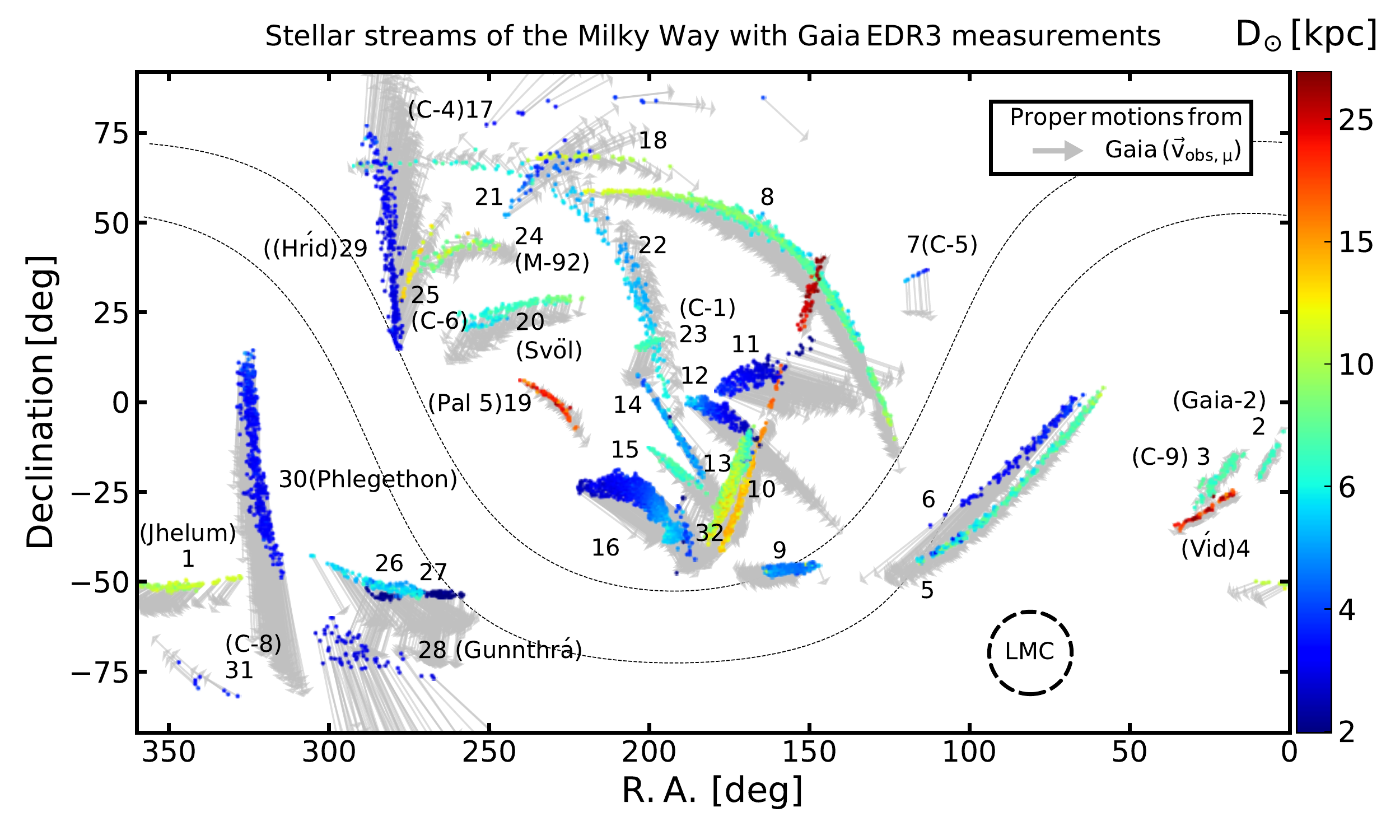}
\end{center}
\vspace{-0.8cm}
\caption{Distribution of stream stars in the Milky Way halo with \Gaia EDR3 measurements. These sources have a $>7\sigma$ likelihood of being stream members according to the {\tt STREAMFINDER} algorithm. The stars are shown in the observed equitorial coordinate. The gray arrows represent the corresponding proper motions of every star (in the same coordinate). The stars are color coded as per their Heliocentric distances. Every stream is tagged with a unique number, and $N=\nstreams$ streams that we eventually use to measure the Sun's motion are labelled by their names. This entire stream catalogue serves as the principal data for our analysis.}
\label{fig:Fig_all_streams_from_SF}
\end{figure*}

At this juncture, it is important to discuss a few points in regard to our method. First, the key assumption of our technique is that the vector $\vperp$ arises entirely due to the reflex motion of the Sun. However, in reality, there could be additional sources that may also contribute to $\vperp$. For instance, if the Milky Way halo is non-spherical (e.g., flattened perpendicular to the disk), then the precession of the angular momentum vector of a stream can contribute to $\vperp$. Nevertheless, even if the halo is flattened by $20\%$, this contribution will be relatively small ($\sim 4\%$ of the observed proper motion, \citetalias{Malhan2017}), so this effect should not bias our measurement of the Sun's velocity. Moreover, the precession corrections will tend to cancel out on average since we analyse multiple streams that move along different trajectories, and whose orbital planes have different inclinations. Another possible source that could contribute to $\vperp$ is if the stream is perturbed due to its past dynamical interaction with the baryonic structures in our Galaxy, namely, the rotating bar, the spiral arms and the Giant Molecular clouds \citep{Amorisco2016, Hattori2016, PriceWhelan2016, Pearson2017}. Such perturbations are known to be very specific to the nature of the stream's orbit, and therefore, streams that would have undergone such perturbative effects are expected to be quite rare\footnote{The Pal~5 stream is the only stream in our data sample that is suspected to have been affected by the bar. Beyond that an example, no stream is yet confirmed to have been perturbed by the Milky Way disk or Giant Molecular clouds.}. Streams can also get perturbed by the gravitational pull of the massive Magellanic Clouds. In the event of such a massive perturbation, the ``kick'' received by the stream stars can cause the stream's velocity vector to become misaligned with the original direction of the stream's motion \citep{Erkal2019, Vasiliev2020}, which may contribute to $\vperp$. However, it has been suggested that the extent of this perturbation should be restricted only to the outer halo regions ($\simgt 30-40\kpc$, e.g., \citealt{Garavito-Camargo2020, Petersen2020}), so the tracers in the inner halo should be impervious to this effect. Although most of the streams in our sample orbit only the inner halo regions ($\simlt 30\kpc$), a few of them do possess large apocentres (e.g., the streams ``Leiptr'' and ``Slidr'', labelled as ``5'' and ``11'' in Fig.~\ref{fig:Fig_all_streams_from_SF}). Whether these large-apocentre streams experienced any perturbation from the Magellanic Clouds is unknown\footnote{This effect is sensitive to the relative orbital motion between the stream and the Magellanic Clouds, and understanding this effect requires proper dynamical modeling.}, although it does not appear to be the case from the visual inspection of their phase-space structure. In summary, we deem that if some of the streams in our data are really perturbed, then the resulting measurement of the Sun's velocity could be biased and very different from those previously reported. However, if the streams are unperturbed, then the resulting Sun's velocity should be a true measurement with respect to the Galactic halo.

\section{Data}\label{sec:data}

Our principal data are $32$ stream-like structures originally detected in the \Gaia Data Release 2 catalog \citep{GaiaDR2_2018_Brown} using the \texttt{STREAMFINDER} algorithm (\citealt{Malhan_SF_2018, Malhan_Ghostly_2018, Ibata_Phlg_2018, Ibata_norse_2019}, Ibata et al., submitted). This stream catalogue comprises a total of $\approx 6000$ sources. Although a number of these streams were known before \Gaia \citep{Grillmair2006GD1_correct, Belokurov2006, Shipp2018}, several were discovered with the \texttt{STREAMFINDER} using \Gaia. A majority of these new streams have been spectroscopically confirmed (Ibata et al., submitted) and our spectroscopic campaign suggests that more than $85\%$ of our sample are bonafide stream members. Note that our data derives from the analysis of \texttt{STREAMFINDER}, and this software detects streams along orbits which are integrated in an assumed Galactic potential model \citep{Malhan_SF_2018, Ibata_norse_2019}. Despite this orbit-based search, the stream detection itself is insensitive to the choice of the potential model (as shown in \citealt{Malhan_SF_2018}). This point is important because we want to highlight that our resulting value of the Sun's velocity, which we measure using these streams, does not depend on any Galactic mass model (not even that used in the \texttt{STREAMFINDER} code). 

We cross-matched this stream catalogue with the recently published ESA/\Gaia EDR3 dataset \citep{GaiaCollaboration2016, GaiaEDR3_Brown_2020, GaiaEDR3_Lindegren_2020, GaiaEDR3_Riello_2020}  (using the VizieR\footnote{\url{https://vizier.u-strasbg.fr/viz-bin/VizieR?-source=I/350&-to=3}} service) in order to take advantage of the new astrometric solutions and new photometry (in the $G, G_{\rm BP}, G_{\rm RP}$ pass-bands). The photometry information is required to compute the photometric distances to the stars (this computation is described below). An ideal scenario would have been to compute distances directly from the \Gaia parallaxes; however, parallaxes for halo stars have large associated uncertainties and their direct use can adversely affect our analysis. Henceforth, all the observed quantities are the \Gaia EDR3 measurements. We extinction-correct the \Gaia EDR3 using the \cite{Schlafly2011} corrections to the \cite{Schlegel1998} extinction maps, assuming the  extinction ratios $A_{\rm G}/A_{\rm V}=0.86117, A_{\rm BP}/A_{\rm V}=1.06126, A_{\rm RP}/A_{\rm V}=  0.64753$, as listed on the web interface of the PARSEC isochrones \citep{Bressan2012}. Henceforth, all magnitudes and colors refer to these extinction-corrected values. All stream stars in our data are shown in Figure~\ref{fig:Fig_all_streams_from_SF}.

To compute the photometric distances of individual stars in a given stream, we make use of the observed color-magnitude diagram (CMD) of the stream, where the color is described as $(G_{\rm BP}- G_{\rm RP})_0$ and $G_0$ is adopted as the magnitude. The task is to compare this CMD with various Single Stellar Population (SSP) models (that are parameterised by Age and [Fe/H]), select a unique template model that suitably represents the observed CMD, and finally use the selected template to compute the distances. Conventionally, a unique template is chosen by visually comparing the observed CMD with the model CMD. However, such a method may fail for us because CMDs of streams generally possess a small spread along the magnitude direction (along $G_0$ in our case) that occurs because streams have non-zero distance gradients. This makes it challenging to find a suitable SSP model simply by ``eye''. To circumvent this issue, we follow a pragmatic approach. 

First, for each stream, we fix the [Fe/H] parameter of the model to a particular value so as to break the [Fe/H]-Age degeneracy. In order to possess realistic [Fe/H] model values, we obtain spectroscopic [Fe/H] measurements from the literature for $14$ of our streams. These streams and their corresponding [Fe/H] measurements are listed in Table~\ref{tab:table1}. A cross-match with the SDSS/Segue \citep{Yanny2009} and LAMOST \citep{Zhao2012} spectroscopic surveys yields spectroscopic metallicities for another $3$ streams. The first of those is the one labelled as ``14'' (Gaia-1) in Figure~\ref{fig:Fig_all_streams_from_SF}, for which we find two members in SDSS/Segue, from which the average metallicity is [Fe/H]=$-1.36$. For stream ``21'' (C-2), we found one member in the SDSS/Segue with [Fe/H]=$-1.82$. Finally, stream ``29'' (Hr\'{\i}d) has two members in the SDSS/Segue and one member in LAMOST, yielding an average metallicity [Fe/H]=$-1.10$. The description of these specific sources is provided in Table~\ref{tab:table2}. The measured line-of-sight velocities of all of these sources are coincident with the orbital solutions as predicted by the \texttt{STREAMFINDER}, and this confirms their membership spectroscopically. Thus, for a given stream, we use the average of the [Fe/H] measurements of their members as the model [Fe/H] value. However, for the remaining $15$ streams with no spectroscopic information, we fix their model [Fe/H] values using the template model solution from the \texttt{STREAMFINDER} algorithm. 

\begin{figure*}
\begin{center}
\includegraphics[width=\hsize]{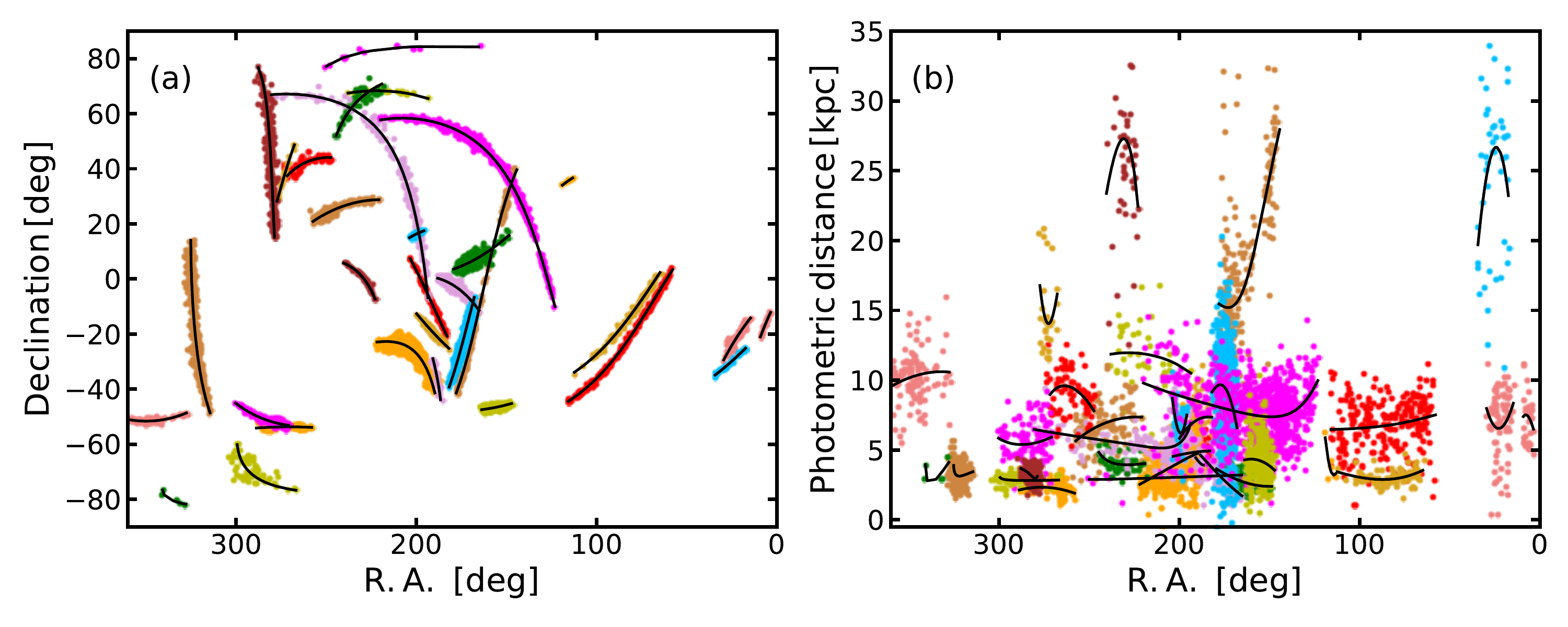}
\end{center}
\vspace{-0.8cm}
\caption{Polynomial fits to the stellar streams in the position space and the distance space. {\bf a:} The observed positions of stream stars. The black curves are the best fit polynomial functions obtained using equation~\ref{eq:phi2_fit} and then transformed to the standard equitorial coordinate. {\bf b:} The photometric distances of the stream stars (computed using the \Gaia EDR3 photometry). The black curves are the best fit polynomial functions obtained using equation~\ref{eq:dis_fit}. Different streams are denoted with different colors to facilitate easier visual discrimination.}
\label{fig:Fig_all_streams_fit_in_RA-Dec-Dist}
\end{figure*}

With an [Fe/H] value assigned to every stream, we then sample the age of the SSP model from $10$--$13\Gyr$, in bins of $0.5\Gyr$, in order to find a unique SSP model that suitably represents the observed CMD of the stream. The means of selecting the unique SSP model is described in Appendix~\ref{appendix:photo_dis}. Finally, we use this chosen SSP model to compute the photometric distances of stream stars. This computation is also described in Appendix~\ref{appendix:photo_dis}. We calibrate these photometric distances with respect to the mean of the orbital-distance solutions of streams (as found by the \texttt{STREAMFINDER}). The resulting distance trends for all the streams are shown in Figure~\ref{fig:Fig_all_streams_fit_in_RA-Dec-Dist}. It is reassuring to find that these distances have similar gradients as those found by the \texttt{STREAMFINDER} algorithm. Henceforth, all distances refer to our computed photometric distances.

\begin{table}
\centering
\caption{Spectroscopic metallicities ([Fe/H]) of streams. The first column provides the labels of the streams (as per Figure~\ref{fig:Fig_all_streams_from_SF}), the second column gives the name of the streams and the third column provides the [Fe/H] values. For most of the streams, the [Fe/H] values are taken from other studies (which are cited in the reference column). For the remaining of the streams, the mentioned [Fe/H] values are only an average/approximate of the spectroscopic [Fe/H] of the member stars that we confirmed based on their velocity information.}
\label{tab:table1}
\begin{tabular}{|l|l|c|c|}

\hline
\hline
{\bf Label} & {\bf Name} & {\bf [Fe/H]} & {\bf reference} \\
&& [dex]&\\
\hline
\hline

1  & Jhelum & -2.0 & \cite{Ji2020} \\
6  & Gj\"oll & -1.78 & \cite{Ibata_norse_2019}\\
8  & GD-1 & -2.24 & \cite{Malhan2019_Pot} \\
9  & NGC~3201 & -1.59 & \cite{Harris2010} \\
11  & Slidr & -1.8 & \cite{Ibata_norse_2019}\\
12  & Sylgr & -2.8 & \cite{Ibata_norse_2019}\\
&&& \cite{Roederer2019}\\
13  & Ylgr & -1.87 & \cite{Ibata_norse_2019}\\
14  & Gaia-1 & -1.36 & our confirmation\\
16  & Fimbulthul & -1.53 & $\rm{\omega \,Cen's}$ value \citep{Harris2010}\\
18  & Kshir & -1.78 & \cite{Malhan2019_Kshir}\\
19  & Pal~5 & -1.41 & \cite{Harris2010} \\
21  & C-2 & -1.82 & our confirmation\\
22  & Fj\"orm & -2.2 & \cite{Ibata_norse_2019}\\
24  & M~92 & -2.31 & \cite{Harris2010} \\
26  & NGC~6397 & -2.02 & \cite{Harris2010} \\
29  & Hr\'{\i}d & -1.1 & our confirmation\\
30  & Phlegethon & -1.56 & \cite{Ibata_Phlg_2018}\\

\hline
\hline
\end{tabular}
\end{table}
\begin{table*}
\centering
\caption{Stars of different streams with spectroscopic observations from SDSS/Segue and LAMOST surveys. The first two columns provide the labels and the names of the streams. Columns 3-4 list the equitorial coordinates R.A. and Declination, respectively. Columns 5-7  provide the names of the source survey, measured metallicities and measured line-of-sight velocities, respectively. The last column provides the predicted velocities from \texttt{STREAMFINDER}.}
\label{tab:table2}
\begin{tabular}{|l|l|c|c|c|c|c|c|}

\hline
\hline
{\bf Label} & {\bf Name} &  {\bf R.A}. & {\bf Dec.} & {\bf Survey} & {\bf [Fe/H]} & {\bf $v_{\rm los}^{\rm data}$} & {\bf $v_{\rm los}^{\rm model}$} \\
&&[deg] & [deg] & &[dex]& [$\kms$]& $[\kms]$\\
\hline
\hline

14  & Gaia-1  &190.883396 & $-8.787459$ &  SDSS/Segue &  $-1.31\pm 0.04$ & $+210\pm 2$ & $+249$\\
& & 198.506958 & 0.319206 & SDSS/Segue &  $-1.41\pm0.03$ & $+97\pm4$ & $+72$\\

21 & C-2 & 244.585242 & 51.865692  & SDSS/Segue & $-1.82\pm  0.08$ & $-396\pm 4$ & $-390$\\

29 & Hr\'{\i}d  & 280.618793 & 40.989063 & SDSS/Segue & $-1.09\pm 0.04$ & $-221\pm 4$ & $-217$\\
& & 279.898775 & 42.446783 & SDSS/Segue & $-1.16\pm 0.04$ & $-216\pm 3$ & $-216$\\
& & 280.736164 & 46.018614 & LAMOST & $-1.05\pm0.04$ & $-201\pm 6$ & $-207$\\

\hline
\hline
\end{tabular}
\end{table*}
\section{Analysis to measure the Sun's velocity}\label{sec:analysis}

The underlying procedure to measure the Sun's velocity is explained in Section~\ref{sec:Method}. Below, we describe how we practically implement that procedure. 

Our first task is to compute the tangent vector $\vdvec$ at every stellar point, which is required to approximate the stream's orbit on the 2D sky. A crude computation of $\vdvec$, by directly applying equation~\ref{eq:vd} to the observed positions, could result in a series of disorderly vectors. This is because streams are not perfect orbits, and every stream has a non-negligible dispersion in phase-space (as can be seen in Fig.~\ref{fig:Fig_all_streams_from_SF}). Thus, prior to the computation of $\vdvec$, we fit each stream in position space with a smooth and continuous curve (which we use afterwards for systematic calculation of $\vdvec$). To this end, for a given stream, we transform the stellar positions from the observed equatorial coordinates $(\alpha,\delta)$ to a set of rotated celestial coordinates $(\phi_1,\phi_2)$. In this coordinate system, $\phi_1$ corresponds to position on a great circle that is approximately parallel to the stream, and $\phi_2$ is the position perpendicular to the stream. In this rotated frame, we fit the stream stars using a quadratic polynomial of the form 
\begin{equation}\label{eq:phi2_fit}
\phi_2 (\phi_1) = a_1 + b_1 \phi_1 + c_1 \phi^2_1 \, ,
\end{equation}
where $\phi_1$ are the coordinates of the data, and $a_1,b_1,c_1$ the fitting parameters to obtain $\phi_2$. We find that this function provides suitable representations of the streams in position space. These fitted parametric curves, for all the streams, are then transformed back to the equatorial coordinate system. The resulting fits to the streams are shown in Figure~\ref{fig:Fig_all_streams_fit_in_RA-Dec-Dist}. Using these best fit models, we then calculate $\vdvec$ for every star (following eq.~\ref{eq:vd}). In \citetalias{Malhan2017}, where we had used stream models to examine our method, this curve fitting was performed independently for the two tidal arms in a given stream. This was necessary because stream models from N-body simulations generally possess two tidal arms that have slightly different energies and angular momenta (e.g., \citealt{Eyre2011}) and, therefore, we had represented the two tidal arms independently. Also, in \citetalias{Malhan2017}, the progenitors of the stream models had survived, and it was possible to easily distinguish between the two tidal arms. In reality, a majority of the observed streams lack any obvious twin tidal arm features. Moreover, we also lack the knowledge of the present day positions of their progenitors. Specifically in our data, we possess the knowledge of the progenitors for only five streams. These include Fimbulthul (the stream of $\omega$ Cen), the stream of NGC~3201, the stream of Pal~5, the stream of M~92, and the stream of NGC~6397. Even among these, the second tidal arm of Fimbulthul is yet unobserved, as it lies close to the Galactic plane in a high-extinction region of the sky. Overall, this makes it nearly impossible for us to identify the two tidal arms for a majority of streams in our data, and then fit the two arms separately. Nevertheless, several studies in the past have worked under the assumption that low-mass streams can be well approximated as one-dimensional structures (e.g., \citealt{Koposov2010, Newberg2010, Malhan2019_Pot}) and, thus, our implementation to fit the entire stream with a single function is a reasonable one. Moreover, any possible systematics due to this effect should average out as long as we analyse a large number of streams distributed across the sky. For the sake of consistency, we make this approximation for all the streams.

To construct the vector $\vobsmu$, we directly use the observed proper motion measurements (following eq.~\ref{eq:vobsmu}), without making any curve fitting in the proper motion space. Finally, vectors $\vdvec$ and $\vobsmu$ allow us to compute $\vperp$ at every stellar point (using eq.~\ref{eq:perp_component}). This gives the ``data'' values that we use in the likelihood function (eq.~\ref{eq:likelihood} below).

In order to compute the model values of $\vperp$ at every stellar point, we require the 2D positions ($\alpha,\delta$) and distances ($D_{\odot}$) of stars. The 2D positions are adopted directly from the fitted curves (that we obtained above). However, we noticed some spread in the distance for every stream ($\sim 20\%$ on average). Thus, in order to correctly account for the distance gradient present in the streams, we fit the distance to stream stars with smoothly varying functions of the form
\begin{equation}\label{eq:dis_fit}
D_{\odot} (\phi_1) = a_2 + b_2 \phi_1 + c_2 \phi^2_1\, ,
\end{equation}
where $a_2,b_2,c_2$ are the fitting parameters to obtain $D_{\odot}$. The best fit models to the streams are shown in Figure~\ref{fig:Fig_all_streams_fit_in_RA-Dec-Dist}.

Finally, we survey the parameter space of ($\VRsun,\Vphisun,\Vzsun$) using our own Metropolis–Hastings based MCMC algorithm. The log-likelihood is defined as
\begin{equation}\label{eq:likelihood}
\ln \mathcal{L} = \sum_{\rm Data} [-\ln (\sigma_{\mu_{\alpha}} \sigma_{\mu_{\delta}}) +\ln N -\ln D]\,,
\end{equation}
where
\begin{equation}
\begin{aligned}
N &= \prod_{j=1}^2 (1-e^{-R^2_j/2}) \, , \\
D &= \prod_{j=1}^2 R^2_j \, , \\
R_1^2 &= \Big(\dfrac{ {\mu_{\bot, \alpha}^{\rm data}} - {\mu_{\bot, \alpha}^{\rm model}}}{\sigma_{\mu_\alpha}} \Big)^2 \, , \\
R_2^2 &= \Big(\dfrac{ {\mu_{\bot, \delta}^{\rm data}} - {\mu_{\bot, \delta}^{\rm model}}}{\sigma_{\mu_\delta}} \Big)^2 \,.
\end{aligned}
\end{equation}

\noindent Here, $\mu_{\bot, \alpha}^{\rm data}$ and $\mu_{\bot, \delta}^{\rm data}$ are the measured components of $\vperp$ along the R.A. and Declination directions. $\mu_{\bot, \alpha}^{\rm model}$ and $\mu_{\bot, \delta}^{\rm model}$ are the corresponding model predictions. These model values are calculated at the position and the distance of every stream star by assuming the Sun's velocity (that we sample as a part of MCMC). The quantities $\sigma_{\mu_\alpha}$ and $\sigma_{\mu_{\delta}}$ are the convolution of the intrinsic proper motion dispersion of the model together with the observational uncertainties of every star. The reason for avoiding the standard log-likelihood function and adopting the ``conservative formulation'' of \cite{sivia1996data}, is to lower the contribution from outliers that could be contaminating our data\footnote{The success of this modified log-likelihood equation, with respect to streams, has been shown in \cite{Malhan2019_Pot}.}. 

In the first run of our analysis, we used all the streams present in our data. However, we found that the MCMC procedure was constraining $\Vphisun$ to an improbably low value of $\sim10\kms$ and the posterior PDF was not smooth. This implied that our assumption about $\vperp$, that it arises primarily due to the Sun's motion, was perhaps not true for all of these structures. To this end, we follow a surgical approach and remove those specific streams for which the assumption about $\vperp$ may in fact not hold. In a nutshell, we remove from the analysis streams that are not smooth in position space, streams with large apocentre ($\simgt 30\kpc$, as they may have been perturbed by the Magellanic Clouds), streams with small pericentre and prograde orbit (as these may have been dynamically perturbed by the Galactic bar or spiral arms), and, finally, streams produced from massive progenitors.

In the first category, we find streams ``10'' (Orphan) and ``26'' (NGC~6397) in Figure~\ref{fig:Fig_all_streams_from_SF}. These streams have ``twisted'' shapes, and thus, are badly fitted in position space. These twists are a possible signature of perturbations produced in the past. While the perturbation in Orphan owes to its past dynamical interaction with the Magellanic Clouds \citep{Erkal2019}, the plausible perturbation in NGC~6397 stream is a tentative observation that we note here.

From the second category, we conservatively remove streams ``5'' (Leiptr), ``6'' (Gj\"oll), ``11'' (Slidr), ``13'' (Ylgr), ``22'' (Fj\"orm), following result from \cite{Ibata_norse_2019}. Using the same argument, we also remove ``9'' (NGC~3201, \citealt{Palau2020}), ``14'' (Gaia~1) and ``21'' ($C-2$).

As per the third category, streams that we deem could have been affected by the structure in the Milky Way disk include the stream of NGC~6397, since the progenitor globular cluster has a prograde motion and it orbits very close to the disk ($z_{\rm max}\sim 3\kpc, r_{\rm apo}\sim 7\kpc$, \citealt{Kalirai2007}). We further remove stream ``19'' that is known to have been perturbed by the Galactic bar (Pal~5; \citealt{Pearson2017}), and stream ``12'' that is on a prograde motion and possesses a very small pericentre ($r_{\rm peri}\simlt 2.5\kpc$) and therefore could have been perturbed by the bar (Sylgr; \citealt{Ibata_norse_2019}). 

Finally, the last category of streams that we remove are those that are known to be associated with massive progenitors, and have high internal velocity dispersion, and thus, (in the ``dynamical'' sense) are not low-mass streams. These include ``16'' (Fimbulthul, the stream of $\omega$ Cen, \citealt{Ibata2019_wcen}), ``8'' and ``18'' (GD-1 and Kshir, \citealt{Malhan2019_Kshir, Gialluca2020}).

After excising these streams, we again ran the MCMC algorithm with the remaining $N=17$ streams, but found $(\VRsun,\Vphisun,\Vzsun)\sim (19,7,23)\kms$. This strange result was not entirely unexpected as some of the streams in our reduced sample, for which the orbits have not yet been constrained, could still be following the above mentioned criteria. Upon closer investigation, we found that streams ``15'' ($C-3$) and ``27'' ($C-7$) were constraining $\Vphisun$ to very low values (as above). This seems to indicate that these streams may well be perturbed. Future spectroscopic observations will shed some light on this conclusion. For our purpose, we remove these streams from our reduced sample. With this, we finally converge to a set of $N=15$ streams which included ``1'' (Jhelum), ``2'' (Gaia-2), ``3'' ($C-9$), ``4'' (V\'{\i}d), ``7'' ($C-5$), ``17'' ($C-4$), ``20'' (Sv\"ol), ``23'' ($C-1$), ``24'' ($M~92$), ``25'' ($C-6$), ``28'' (Gunnthr\'a), ``29'' (Hr\'{\i}d), ``30'' (Phlegethon), ``31'' ($C-8$), ``32''. The analysis of this set of streams yields the following velocity for the Sun: $(\VRsun, \Vphisun, \Vzsun)=(9.90^{+1.15}_{-1.11}, 244.56^{+1.58}_{-1.56}, 2.91^{+0.98}_{-1.00}) \kms$. We use a right-handed coordinate system, in which positive $\VRsun$ is radially towards the Galactic centre, positive $\Vphisun$ is the rotational velocity in the direction of Galactic rotation, and positive $\Vzsun$ is vertically upwards from the disk.

In order to remain as objective as possible in the construction of our stream sample, we test all other streams from our excised sample and add them back to the main sample if the analysis continues to yield consistent results. All streams that produce a change in the Sun's velocity that is within $3\sigma$ of the initial result with 15 streams are re-integrated into our golden sample, whereas the others are left out. This analysis shows that two additional streams can be brought back into the golden sample, whereas the other ones produce deviant results.

\begin{figure}
\begin{center}
\includegraphics[width=\hsize]{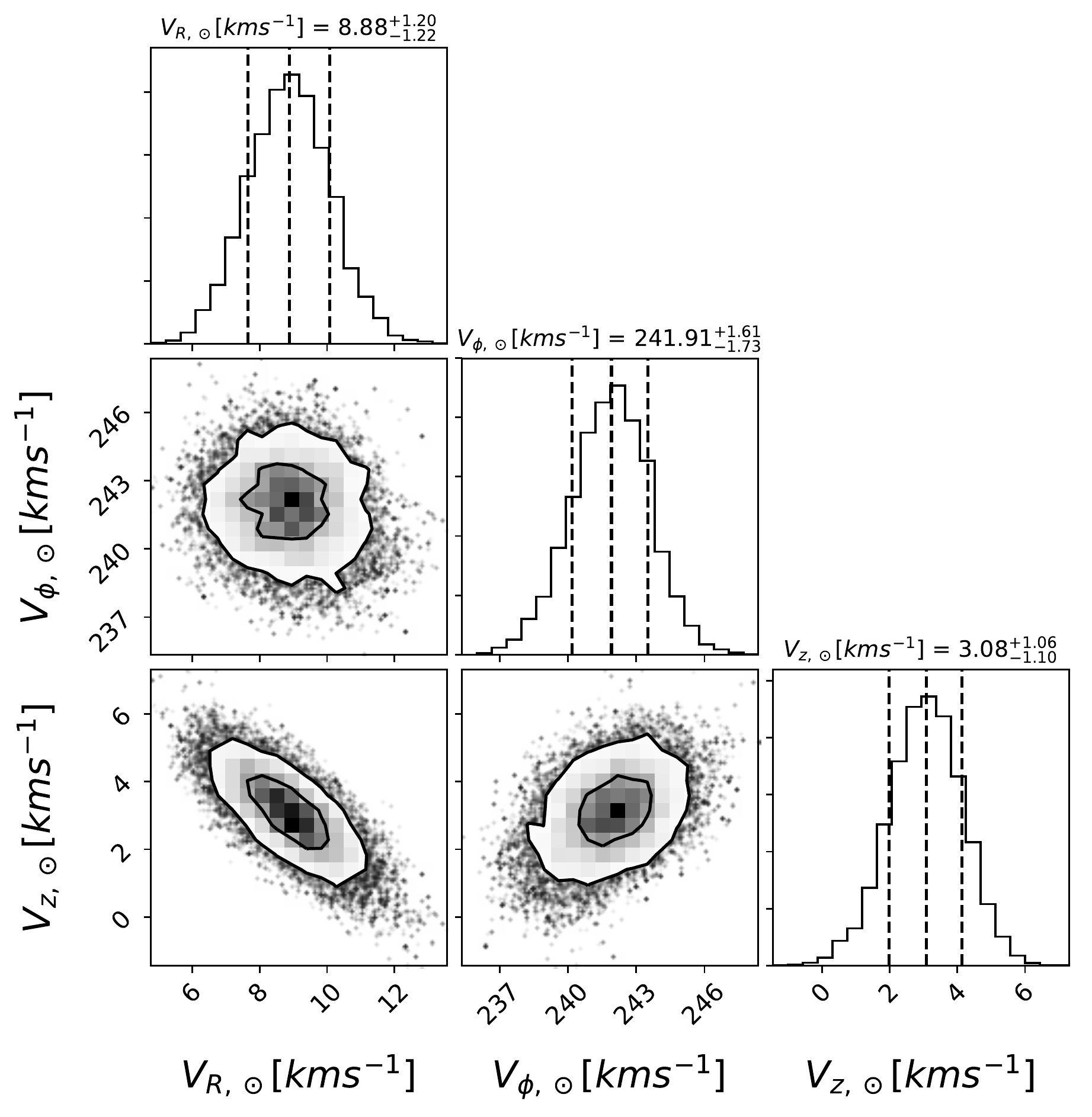}
\end{center}
\vspace{-0.4cm}
\caption{Posteriors PDFs on the $\VRsun,\Vphisun,\Vzsun$ components of the Sun's galactic velocity $\V3Dsun$ obtained by analysing $N=\nstreams$ streams of the Galactic halo. Our coordinate system is such that positive $\VRsun$ is radially towards the Galactic centre, positive $\Vphisun$ is the total rotational velocity in the direction of Galactic rotation, and positive $\Vzsun$ is vertically upwards from the Galactic disk. For the 1D PDFs, the dashed lines correspond to quantiles $(0.16,0.50,0.84)$, and for the 2D PDFs, the contours correspond to $\pm 1\sigma,\pm 2\sigma$ range (assuming Gaussian functions). Our inference shows a strong anti-correlation between $\VRsun$ and $\Vzsun$.}
\label{fig:Corner_Plot_2}
\end{figure}

The results of our final analysis based on this sample of 17 streams are displayed in Figure~\ref{fig:Corner_Plot_2} and yield the following constraint on the Sun's velocity: $(\VRsun,\Vphisun,\Vzsun) = (\VRsunA, \VphisunA, \VzsunA) \kms$, but do note that the components $\VRsun$ and $\Vzsun$ are significantly correlated\footnote{To facilitate future uses of our inferred PDFs, we publish our MCMC chain along with this paper (see Appendix~\ref{appendix:MCMC_chain}).}. Note also that the quoted uncertainties on the velocity components are purely formal. A source of additional systematic error could stem from systematic biases on the distances, which could arise from an potentially incorrect calibration in the photometric distances. However, since we calibrate our photometric distances based on the orbital solutions of streams (as found by the \texttt{STREAMFINDER} algorithm), we expect that this effect should be small.

\begin{figure*}
\begin{center}
\includegraphics[width=\hsize]{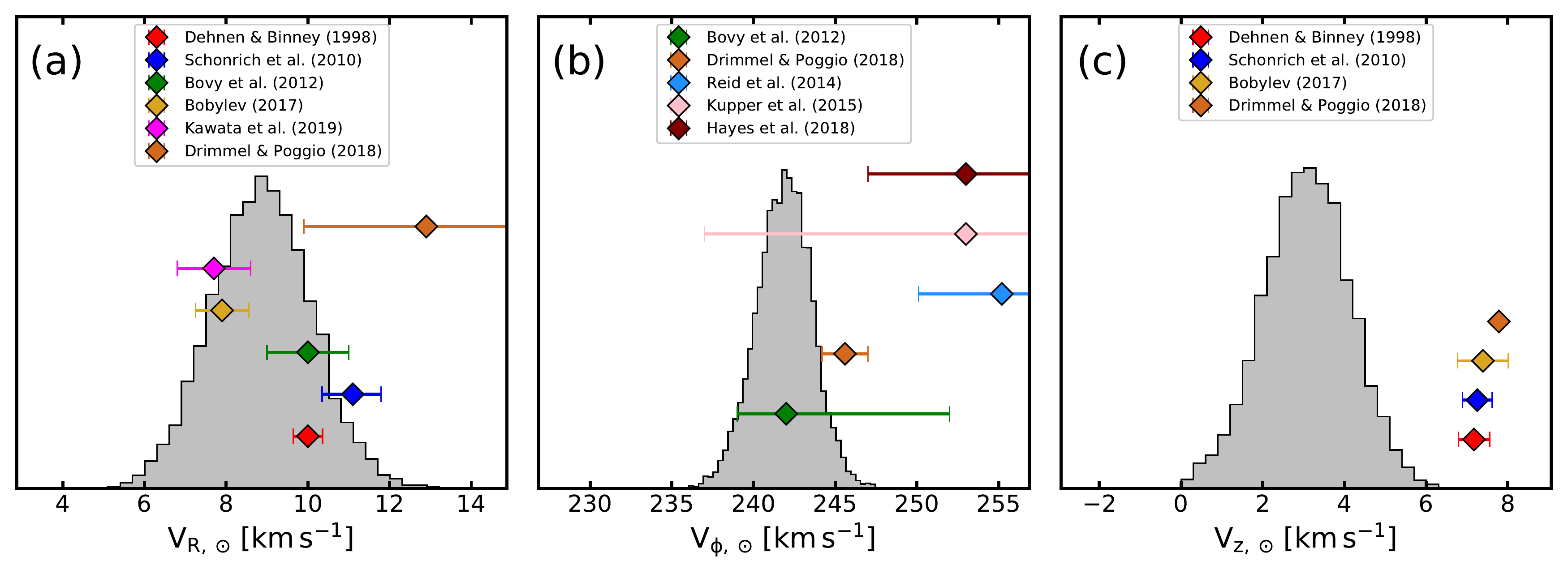}
\end{center}
\vspace{-0.6cm}
\caption{Comparing the $\VRsun,\Vphisun,\Vzsun$ components of the Sun's galactic velocity $\V3Dsun$ between different studies. The gray histograms correspond to the posteriors that we obtained by analysing $N=\nstreams$ streams of the Galactic halo. The colored markers correspond to the measurements from other studies that analysed the tracers located either at the Galactic center \citep{Drimmel_2018}, or in the Galactic disk \citep{Bovy2012, Reid2014} or in the solar neighbourhood \citep{DehnenBinney1998_Sun, Schonrich2010, Bobylev2017, Kawata2019}. \citet{Kupper2015} and \citet{Hayes2018} analysed the Pal5 stream and the Sagittarius stream, respectively, and constrained $\Vphisun$ assuming a Galactic potential model. While our measurements of $\VRsun,\Vphisun$ are nearly consistent with the previous studies, however, we find our value of $\Vzsun$ is $\sim 5\kms$ smaller in comparison with previous measurements. This implies that the Sun's velocity depends subtly on the choice of the reference frame (which in our case is the `Galactic halo' as defined by the stellar streams).}
\label{fig:Comparison_Vsun}
\end{figure*}
\begin{table*}
\centering
\caption{The Sun's velocity in the reference frames of the Galactic Centre, the Galactic disk and the Galactic Halo. To obtain the Sun's velocity relative to the Galactic centre, we compute $\Vphisun,\Vzsun$ using proper motion of Sagittarius $A^{*}$ (along the Galactic plane=$6.411 \pm 0.008 \masyr$, perpendicular to the Galactic plane=$0.219\pm0.007 \masyr$, \citealt{Reid2020}) and the Sun's galactic distance ($8.178 \pm 0.013\kpc$, \citealt{GravityCollaboration2019}), and for $\VRsun$ we adopt the value of $vz_0$ from \citet{GravityCollaboration2019}. For the Sun's velocity in the Galactic disk, we take $\Vphisun$ from \citet{Reid2014}, and $\VRsun,\Vzsun$ from \citet{Schonrich2010}. The fourth column provides the Sun's speed that we compute by taking the norm of the velocity vector.}
\label{tab:table3}
\begin{tabular}{|l|l|c|c|c|c|c|c|}

\hline
\hline
{\bf Reference frame} & $\VRsun$ &  $\Vphisun$ & $\Vzsun$ & $||\V3Dsun||$ & {\bf reference} \\
& $[\kms]$ & $[\kms]$& $[\kms]$ & $[\kms]$ &\\
\hline
\hline

Galactic Centre & $14.1\pm 1.7$ & $248.5\pm0.5$ & $8.49\pm0.27$ & $249.1\pm 0.5$ & \cite{Reid2020}\\
& & & & & \cite{GravityCollaboration2019}\\
& & & & & \cite{Schonrich2010}\\

Galactic Disk & $11.1^{+0.69}_{-0.75}$ & $255.2\pm 5.1$ & $7.25^{+0.37}_{-0.36}$ & $256\pm 5$ & \cite{Schonrich2010}\\
& & & & & \cite{Reid2014}\\

Galactic Halo & $\VRsunA$ & $\VphisunA$ & $\VzsunA$ & $242.1\pm 1.7$ & This work\\

\hline
\hline
\end{tabular}
\end{table*}

\vfill\eject
\section{Conclusions and Discussion}\label{sec:conclusion_and_discussion}

Traditionally, the Sun's velocity is measured using only those tracers that are located either at the Galactic centre or in the Galactic disk. In this contribution, we measured the Sun's velocity with respect to the stellar streams that orbit the Galactic halo, with astrometric parameters derived from the recently published ESA/\Gaia EDR3 dataset. This is the first time that all three components of the Sun's velocity, $\V3Dsun$, are measured with respect to the Galactic halo. Our method relies on a very basic behaviour of low-mass streams: that the proper motion of stream stars should be closely directed along the stream structure itself in the rest frame of the Galaxy, and that the observed perpendicular motion of stars arises (primarily) from the Sun's own reflex motion. Building on this principle, we employed a simple geometrical procedure on a sample of $N=\nstreams$ streams and measured the Sun's velocity as $(\VRsun,\Vphisun,\Vzsun) = (\VRsunA,\VphisunA,\VzsunA)\kms$. Here, positive $\VRsun$ implies radially towards the Galactic centre, positive $\Vphisun$ implies the total rotational velocity in the direction of Galactic rotation and positive $\Vzsun$ implies vertically upwards from the Galactic disk. We highlight that this measurement is independent of any Galactic potential model, and it is also uncorrelated with the Sun's galactic distance ($\Rsun$). We now discuss how our measurement compares with those previously obtained using different techniques (a summary of this comparison is also shown in Fig.~\ref{fig:Comparison_Vsun}).

The $\VRsun$ component of the Sun's motion has been previously measured in the range of $\approx 8-11\kms$ (radially towards the Galactic centre), based on studies that have analysed tracers in the Galactic disk or in the solar neighbourhood (e.g., \citealt{DehnenBinney1998_Sun, Schonrich2010, Bovy2012,  Bobylev2017, Kawata2019}). \cite{Drimmel_2018} calculated $\VRsun=12.9\pm3.0\kms$ using the study of \cite{Gravity_Collaboration2018}. Our measurement of $\VRsun$ is in good agreement with these values. 

Previous measurements of $\Vzsun$ agree on the value of $\approx 7\kms$ (vertically upwards from the Galactic disk, e.g., \citealt{DehnenBinney1998_Sun, Schonrich2010, Bobylev2017}). These studies analysed the dynamics of tracers in the solar neighbourhood. Alternatively, $\Vzsun$ can be computed by multiplying the value of $\Rsun$ with the proper motion of Sagittarius $A^{*}$ perpendicular to the Galactic plane, from which one obtains $\Vzsun=8.49\pm 0.27\kms$ (see Table~\ref{tab:table3}). This value of $\Vzsun$ can be interpreted as being a measurement with respect to the Galactic center. Our measurement of $\Vzsun$ is slightly but significantly lower than these values. If our hypotheses hold, this implies that the vertical velocity of the Sun with respect to the halo is $\sim 4-5\kms$ smaller than that with respect to the Galactic centre/disk.

Estimates of the Sun's motion in the direction of Galactic rotation, $\Vphisun$,  generally range from $\sim 220-260\kms$. However, many of the recent estimates specifically favour a higher value of $\sim 240-260\kms$ (e.g., \citealt{Bovy2012, Reid2014, Kupper2015, Drimmel_2018, Hayes2018}). Our measurement of $\Vphisun$ is in good agreement with these values. However, $\Vphisun$ can be alternatively computed by multiplying the value of $\Rsun$ with the proper motion of Sagittarius $A^{*}$ along the Galactic plane, for which one obtains $\Vphisun=248.5 \pm 0.5 \kms$ (see Table~\ref{tab:table3}). Our measurement of $\Vphisun$ is slightly, but significantly, lower than this particular value.

Here, we measured the Sun's velocity with respect to the Galactic halo, finding a very similar value compared to previous work that used the Galactic disk and Galactic center as references. As such, our measurement provides a useful external check on other methods of measuring the velocity of the Sun. However, it also shows that the Galaxy within the Solar circle is at rest, or almost at rest, with respect to the dark matter halo over the region traversed by the stream sample ($\sim 30\kpc$ radius). It will be interesting to fold in additional new streams into our analysis as new data become available. If the discrepancy in the $z$-direction motion holds up, it would provide evidence for differential motion between the inner and outer Galaxy, which might arise, for instance, from an external perturbation.

\section*{ACKNOWLEDGEMENTS}

KM acknowledges support  from the $\rm{Vetenskapsr\mathring{a}de}$t (Swedish Research Council) through contract No. 638-2013-8993 and the Oskar Klein Centre for Cosmoparticle Physics. RI and NM acknowledge funding from the Agence Nationale de la Recherche (ANR-18-CE31-0017) and from the European Research Council (ERC) under the European Unions Horizon 2020 research and innovation programme (grant agreement No. 834148). 

This research has made use of the VizieR catalogue access tool, CDS, Strasbourg, France (DOI : 10.26093/cds/vizier). The original description of the VizieR service was published in 2000, A\& AS 143, 23.

This work has made use of data from the European Space Agency (ESA) mission {\it Gaia} (\url{https://www.cosmos.esa.int/gaia}), processed by the {\it Gaia} Data Processing and Analysis Consortium (DPAC,
\url{https://www.cosmos.esa.int/web/gaia/dpac/consortium}). Funding for the DPAC has been provided by national institutions, in particular the institutions participating in the {\it Gaia} Multilateral Agreement. 
 
Funding for SDSS-III has been provided by the Alfred P. Sloan Foundation, the Participating Institutions, the National Science Foundation, and the U.S. Department of Energy Office of Science. The SDSS-III web site is http://www.sdss3.org/.

SDSS-III is managed by the Astrophysical Research Consortium for the Participating Institutions of the SDSS-III Collaboration including the University of Arizona, the Brazilian Participation Group, Brookhaven National Laboratory, Carnegie Mellon University, University of Florida, the French Participation Group, the German Participation Group, Harvard University, the Instituto de Astrofisica de Canarias, the Michigan State/Notre Dame/JINA Participation Group, Johns Hopkins University, Lawrence Berkeley National Laboratory, Max Planck Institute for Astrophysics, Max Planck Institute for Extraterrestrial Physics, New Mexico State University, New York University, Ohio State University, Pennsylvania State University, University of Portsmouth, Princeton University, the Spanish Participation Group, University of Tokyo, University of Utah, Vanderbilt University, University of Virginia, University of Washington, and Yale University.

Guoshoujing Telescope (the Large Sky Area Multi-Object Fiber Spectroscopic Telescope LAMOST) is a National Major Scientific Project built by the Chinese Academy of Sciences. Funding for the project has been provided by the National Development and Reform Commission. LAMOST is operated and managed by the National Astronomical Observatories, Chinese Academy of Sciences.

\bibliographystyle{mnras}
\bibliography{ref1} % if your bibtex file is called example.bib

\begin{thebibliography}{}
\makeatletter
\relax
\def\mn@urlcharsother{\let\do\@makeother \do\$\do\&\do\#\do\^\do\_\do\%\do\~}
\def\mn@doi{\begingroup\mn@urlcharsother \@ifnextchar [ {\mn@doi@}
  {\mn@doi@[]}}
\def\mn@doi@[#1]#2{\def\@tempa{#1}\ifx\@tempa\@empty \href
  {http://dx.doi.org/#2} {doi:#2}\else \href {http://dx.doi.org/#2} {#1}\fi
  \endgroup}
\def\mn@eprint#1#2{\mn@eprint@#1:#2::\@nil}
\def\mn@eprint@arXiv#1{\href {http://arxiv.org/abs/#1} {{\tt arXiv:#1}}}
\def\mn@eprint@dblp#1{\href {http://dblp.uni-trier.de/rec/bibtex/#1.xml}
  {dblp:#1}}
\def\mn@eprint@#1:#2:#3:#4\@nil{\def\@tempa {#1}\def\@tempb {#2}\def\@tempc
  {#3}\ifx \@tempc \@empty \let \@tempc \@tempb \let \@tempb \@tempa \fi \ifx
  \@tempb \@empty \def\@tempb {arXiv}\fi \@ifundefined
  {mn@eprint@\@tempb}{\@tempb:\@tempc}{\expandafter \expandafter \csname
  mn@eprint@\@tempb\endcsname \expandafter{\@tempc}}}

\bibitem[\protect\citeauthoryear{{Amorisco}, {G{\'o}mez}, {Vegetti}  \&
  {White}}{{Amorisco} et~al.}{2016}]{Amorisco2016}
{Amorisco} N.~C.,  {G{\'o}mez} F.~A.,  {Vegetti} S.,   {White} S. D.~M.,  2016,
  \mn@doi [\mnras] {10.1093/mnrasl/slw148}, \href
  {https://ui.adsabs.harvard.edu/abs/2016MNRAS.463L..17A} {463, L17}

\bibitem[\protect\citeauthoryear{{Antoja} et~al.,}{{Antoja}
  et~al.}{2018}]{Antoja2018}
{Antoja} T.,  et~al., 2018, \mn@doi [\nat] {10.1038/s41586-018-0510-7}, \href
  {https://ui.adsabs.harvard.edu/abs/2018Natur.561..360A} {561, 360}

\bibitem[\protect\citeauthoryear{{Belokurov} et~al.,}{{Belokurov}
  et~al.}{2006}]{Belokurov2006}
{Belokurov} V.,  et~al., 2006, \mn@doi [\apjl] {10.1086/504797}, \href
  {https://ui.adsabs.harvard.edu/abs/2006ApJ...642L.137B} {642, L137}

\bibitem[\protect\citeauthoryear{{Bobylev}}{{Bobylev}}{2017}]{Bobylev2017}
{Bobylev} V.~V.,  2017, \mn@doi [Astronomy Letters]
  {10.1134/S106377371703001X}, \href
  {https://ui.adsabs.harvard.edu/abs/2017AstL...43..152B} {43, 152}

\bibitem[\protect\citeauthoryear{{Bovy}}{{Bovy}}{2015}]{Bovy2015}
{Bovy} J.,  2015, \mn@doi [\apjs] {10.1088/0067-0049/216/2/29}, \href
  {https://ui.adsabs.harvard.edu/abs/2015ApJS..216...29B} {216, 29}

\bibitem[\protect\citeauthoryear{{Bovy}, {Hogg}  \& {Rix}}{{Bovy}
  et~al.}{2009}]{Bovy2009}
{Bovy} J.,  {Hogg} D.~W.,   {Rix} H.-W.,  2009, \mn@doi [\apj]
  {10.1088/0004-637X/704/2/1704}, \href
  {https://ui.adsabs.harvard.edu/abs/2009ApJ...704.1704B} {704, 1704}

\bibitem[\protect\citeauthoryear{{Bovy} et~al.,}{{Bovy}
  et~al.}{2012}]{Bovy2012}
{Bovy} J.,  et~al., 2012, \mn@doi [\apj] {10.1088/0004-637X/759/2/131}, \href
  {https://ui.adsabs.harvard.edu/abs/2012ApJ...759..131B} {759, 131}

\bibitem[\protect\citeauthoryear{{Bowden}, {Belokurov}  \& {Evans}}{{Bowden}
  et~al.}{2015}]{Bowden2015}
{Bowden} A.,  {Belokurov} V.,   {Evans} N.~W.,  2015, \mn@doi [\mnras]
  {10.1093/mnras/stv285}, \href
  {https://ui.adsabs.harvard.edu/abs/2015MNRAS.449.1391B} {449, 1391}

\bibitem[\protect\citeauthoryear{{Bressan}, {Marigo}, {Girardi}, {Salasnich},
  {Dal Cero}, {Rubele}  \& {Nanni}}{{Bressan} et~al.}{2012}]{Bressan2012}
{Bressan} A.,  {Marigo} P.,  {Girardi} L.,  {Salasnich} B.,  {Dal Cero} C.,
  {Rubele} S.,   {Nanni} A.,  2012, \mn@doi [\mnras]
  {10.1111/j.1365-2966.2012.21948.x}, \href
  {https://ui.adsabs.harvard.edu/abs/2012MNRAS.427..127B} {427, 127}

\bibitem[\protect\citeauthoryear{{Dehnen} \& {Binney}}{{Dehnen} \&
  {Binney}}{1998a}]{DehnenBinney1998}
{Dehnen} W.,  {Binney} J.,  1998a, \mn@doi [\mnras]
  {10.1046/j.1365-8711.1998.01282.x}, \href
  {https://ui.adsabs.harvard.edu/abs/1998MNRAS.294..429D} {294, 429}

\bibitem[\protect\citeauthoryear{{Dehnen} \& {Binney}}{{Dehnen} \&
  {Binney}}{1998b}]{DehnenBinney1998_Sun}
{Dehnen} W.,  {Binney} J.~J.,  1998b, \mn@doi [\mnras]
  {10.1046/j.1365-8711.1998.01600.x}, \href
  {https://ui.adsabs.harvard.edu/abs/1998MNRAS.298..387D} {298, 387}

\bibitem[\protect\citeauthoryear{Dehnen, Odenkirchen, Grebel  \& Rix}{Dehnen
  et~al.}{2004}]{Dehnen2004}
Dehnen W.,  Odenkirchen M.,  Grebel E.~K.,   Rix H.-W.,  2004, \mn@doi [The
  Astronomical Journal] {10.1086/383214}, 127, 2753

\bibitem[\protect\citeauthoryear{Drimmel \& Poggio}{Drimmel \&
  Poggio}{2018}]{Drimmel_2018}
Drimmel R.,  Poggio E.,  2018, \mn@doi [Research Notes of the {AAS}]
  {10.3847/2515-5172/aaef8b}, 2, 210

\bibitem[\protect\citeauthoryear{{Eilers}, {Hogg}, {Rix}  \& {Ness}}{{Eilers}
  et~al.}{2019}]{Eilers2019}
{Eilers} A.-C.,  {Hogg} D.~W.,  {Rix} H.-W.,   {Ness} M.~K.,  2019, \mn@doi
  [\apj] {10.3847/1538-4357/aaf648}, \href
  {https://ui.adsabs.harvard.edu/abs/2019ApJ...871..120E} {871, 120}

\bibitem[\protect\citeauthoryear{{Erkal} et~al.,}{{Erkal}
  et~al.}{2019}]{Erkal2019}
{Erkal} D.,  et~al., 2019, \mn@doi [\mnras] {10.1093/mnras/stz1371}, \href
  {https://ui.adsabs.harvard.edu/abs/2019MNRAS.487.2685E} {487, 2685}

\bibitem[\protect\citeauthoryear{{Eyre} \& {Binney}}{{Eyre} \&
  {Binney}}{2011}]{Eyre2011}
{Eyre} A.,  {Binney} J.,  2011, \mn@doi [\mnras]
  {10.1111/j.1365-2966.2011.18270.x}, \href
  {https://ui.adsabs.harvard.edu/abs/2011MNRAS.413.1852E} {413, 1852}

\bibitem[\protect\citeauthoryear{{Freese}, {Lisanti}  \& {Savage}}{{Freese}
  et~al.}{2013}]{Freese2013}
{Freese} K.,  {Lisanti} M.,   {Savage} C.,  2013, \mn@doi [Reviews of Modern
  Physics] {10.1103/RevModPhys.85.1561}, \href
  {https://ui.adsabs.harvard.edu/abs/2013RvMP...85.1561F} {85, 1561}

\bibitem[\protect\citeauthoryear{{Gaia Collaboration} et~al.,}{{Gaia
  Collaboration} et~al.}{2016}]{GaiaCollaboration2016}
{Gaia Collaboration} et~al., 2016, \mn@doi [\aap]
  {10.1051/0004-6361/201629272}, \href
  {https://ui.adsabs.harvard.edu/abs/2016A&A...595A...1G} {595, A1}

\bibitem[\protect\citeauthoryear{{Gaia Collaboration}, {Brown, A. G. A.},
  {Vallenari, A.}, {Prusti, T.}, {de Bruijne, J. H. J.}  \& {et al.}}{{Gaia
  Collaboration} et~al.}{2018}]{GaiaDR2_2018_Brown}
{Gaia Collaboration} {Brown, A. G. A.} {Vallenari, A.} {Prusti, T.} {de
  Bruijne, J. H. J.}  {et al.} 2018, \mn@doi [A\&A]
  {10.1051/0004-6361/201833051}

\bibitem[\protect\citeauthoryear{{Gaia Collaboration}, {Brown, Anthony G.A.},
  {Vallenari, A.}, {Prusti, T.}  \& {de Bruijne, J. H.J.}}{{Gaia Collaboration}
  et~al.}{2020}]{GaiaEDR3_Brown_2020}
{Gaia Collaboration} {Brown, Anthony G.A.} {Vallenari, A.} {Prusti, T.}  {de
  Bruijne, J. H.J.} 2020, \mn@doi [A\&A] {10.1051/0004-6361/202039657}

\bibitem[\protect\citeauthoryear{{Garavito-Camargo}, {Besla}, {Laporte},
  {Price-Whelan}, {Cunningham}, {Johnston}, {Weinberg}  \&
  {Gomez}}{{Garavito-Camargo} et~al.}{2020}]{Garavito-Camargo2020}
{Garavito-Camargo} N.,  {Besla} G.,  {Laporte} C. F.~P.,  {Price-Whelan} A.~M.,
   {Cunningham} E.~C.,  {Johnston} K.~V.,  {Weinberg} M.~D.,   {Gomez} F.~A.,
  2020, arXiv e-prints, \href
  {https://ui.adsabs.harvard.edu/abs/2020arXiv201000816G} {p. arXiv:2010.00816}

\bibitem[\protect\citeauthoryear{{Ghez} et~al.,}{{Ghez}
  et~al.}{2008}]{Ghez2008}
{Ghez} A.~M.,  et~al., 2008, \mn@doi [\apj] {10.1086/592738}, \href
  {https://ui.adsabs.harvard.edu/abs/2008ApJ...689.1044G} {689, 1044}

\bibitem[\protect\citeauthoryear{{Gialluca}, {Naidu}  \& {Bonaca}}{{Gialluca}
  et~al.}{2020}]{Gialluca2020}
{Gialluca} M.~T.,  {Naidu} R.~P.,   {Bonaca} A.,  2020, arXiv e-prints, \href
  {https://ui.adsabs.harvard.edu/abs/2020arXiv201112963G} {p. arXiv:2011.12963}

\bibitem[\protect\citeauthoryear{{Gravity Collaboration} et~al.,}{{Gravity
  Collaboration} et~al.}{2018}]{Gravity_Collaboration2018}
{Gravity Collaboration} et~al., 2018, \mn@doi [\aap]
  {10.1051/0004-6361/201833718}, \href
  {https://ui.adsabs.harvard.edu/abs/2018A&A...615L..15G} {615, L15}

\bibitem[\protect\citeauthoryear{{Gravity Collaboration} et~al.,}{{Gravity
  Collaboration} et~al.}{2019}]{GravityCollaboration2019}
{Gravity Collaboration} et~al., 2019, \mn@doi [\aap]
  {10.1051/0004-6361/201935656}, \href
  {https://ui.adsabs.harvard.edu/abs/2019A&A...625L..10G} {625, L10}

\bibitem[\protect\citeauthoryear{{Grillmair} \& {Dionatos}}{{Grillmair} \&
  {Dionatos}}{2006}]{Grillmair2006GD1_correct}
{Grillmair} C.~J.,  {Dionatos} O.,  2006, \mn@doi [\apjl] {10.1086/505111},
  \href {http://adsabs.harvard.edu/abs/2006ApJ...643L..17G} {643, L17}

\bibitem[\protect\citeauthoryear{{Harris}}{{Harris}}{2010}]{Harris2010}
{Harris} W.~E.,  2010, arXiv e-prints, \href
  {https://ui.adsabs.harvard.edu/abs/2010arXiv1012.3224H} {p. arXiv:1012.3224}

\bibitem[\protect\citeauthoryear{{Hattori}, {Erkal}  \& {Sanders}}{{Hattori}
  et~al.}{2016}]{Hattori2016}
{Hattori} K.,  {Erkal} D.,   {Sanders} J.~L.,  2016, \mn@doi [\mnras]
  {10.1093/mnras/stw1006}, \href
  {https://ui.adsabs.harvard.edu/abs/2016MNRAS.460..497H} {460, 497}

\bibitem[\protect\citeauthoryear{{Hayes}, {Law}  \& {Majewski}}{{Hayes}
  et~al.}{2018}]{Hayes2018}
{Hayes} C.~R.,  {Law} D.~R.,   {Majewski} S.~R.,  2018, \mn@doi [\apjl]
  {10.3847/2041-8213/aae9dd}, \href
  {https://ui.adsabs.harvard.edu/abs/2018ApJ...867L..20H} {867, L20}

\bibitem[\protect\citeauthoryear{{Ibata}, {Malhan}, {Martin}  \&
  {Starkenburg}}{{Ibata} et~al.}{2018}]{Ibata_Phlg_2018}
{Ibata} R.~A.,  {Malhan} K.,  {Martin} N.~F.,   {Starkenburg} E.,  2018,
  \mn@doi [\apj] {10.3847/1538-4357/aadba3}, \href
  {https://ui.adsabs.harvard.edu/abs/2018ApJ...865...85I} {865, 85}

\bibitem[\protect\citeauthoryear{{Ibata}, {Bellazzini}, {Malhan}, {Martin}  \&
  {Bianchini}}{{Ibata} et~al.}{2019a}]{Ibata2019_wcen}
{Ibata} R.~A.,  {Bellazzini} M.,  {Malhan} K.,  {Martin} N.,   {Bianchini} P.,
  2019a, \mn@doi [Nature Astronomy] {10.1038/s41550-019-0751-x}, \href
  {https://ui.adsabs.harvard.edu/abs/2019NatAs...3..667I} {3, 667}

\bibitem[\protect\citeauthoryear{{Ibata}, {Malhan}  \& {Martin}}{{Ibata}
  et~al.}{2019b}]{Ibata_norse_2019}
{Ibata} R.~A.,  {Malhan} K.,   {Martin} N.~F.,  2019b, \mn@doi [\apj]
  {10.3847/1538-4357/ab0080}, \href
  {https://ui.adsabs.harvard.edu/abs/2019ApJ...872..152I} {872, 152}

\bibitem[\protect\citeauthoryear{{Ji} et~al.,}{{Ji} et~al.}{2020}]{Ji2020}
{Ji} A.~P.,  et~al., 2020, \mn@doi [\aj] {10.3847/1538-3881/abacb6}, \href
  {https://ui.adsabs.harvard.edu/abs/2020AJ....160..181J} {160, 181}

\bibitem[\protect\citeauthoryear{{Kalirai} et~al.,}{{Kalirai}
  et~al.}{2007}]{Kalirai2007}
{Kalirai} J.~S.,  et~al., 2007, \mn@doi [\apjl] {10.1086/513102}, \href
  {https://ui.adsabs.harvard.edu/abs/2007ApJ...657L..93K} {657, L93}

\bibitem[\protect\citeauthoryear{{Kawata}, {Bovy}, {Matsunaga}  \&
  {Baba}}{{Kawata} et~al.}{2019}]{Kawata2019}
{Kawata} D.,  {Bovy} J.,  {Matsunaga} N.,   {Baba} J.,  2019, \mn@doi [\mnras]
  {10.1093/mnras/sty2623}, \href
  {https://ui.adsabs.harvard.edu/abs/2019MNRAS.482...40K} {482, 40}

\bibitem[\protect\citeauthoryear{{Koposov}, {Rix}  \& {Hogg}}{{Koposov}
  et~al.}{2010}]{Koposov2010}
{Koposov} S.~E.,  {Rix} H.-W.,   {Hogg} D.~W.,  2010, \mn@doi [\apj]
  {10.1088/0004-637X/712/1/260}, \href
  {https://ui.adsabs.harvard.edu/abs/2010ApJ...712..260K} {712, 260}

\bibitem[\protect\citeauthoryear{{K{\"u}pper}, {Balbinot}, {Bonaca},
  {Johnston}, {Hogg}, {Kroupa}  \& {Santiago}}{{K{\"u}pper}
  et~al.}{2015}]{Kupper2015}
{K{\"u}pper} A. H.~W.,  {Balbinot} E.,  {Bonaca} A.,  {Johnston} K.~V.,  {Hogg}
  D.~W.,  {Kroupa} P.,   {Santiago} B.~X.,  2015, \mn@doi [\apj]
  {10.1088/0004-637X/803/2/80}, \href
  {https://ui.adsabs.harvard.edu/abs/2015ApJ...803...80K} {803, 80}

\bibitem[\protect\citeauthoryear{{Lindegren, Lennart} et~al.,}{{Lindegren,
  Lennart} et~al.}{2020}]{GaiaEDR3_Lindegren_2020}
{Lindegren, Lennart} et~al., 2020, \mn@doi [A\&A] {10.1051/0004-6361/202039709}

\bibitem[\protect\citeauthoryear{{Majewski}, {Law}, {Polak}  \&
  {Patterson}}{{Majewski} et~al.}{2006}]{Majewski2006}
{Majewski} S.~R.,  {Law} D.~R.,  {Polak} A.~A.,   {Patterson} R.~J.,  2006,
  \mn@doi [\apjl] {10.1086/500195}, \href
  {https://ui.adsabs.harvard.edu/abs/2006ApJ...637L..25M} {637, L25}

\bibitem[\protect\citeauthoryear{{Malhan} \& {Ibata}}{{Malhan} \&
  {Ibata}}{2017}]{Malhan2017}
{Malhan} K.,  {Ibata} R.~A.,  2017, \mn@doi [\mnras] {10.1093/mnras/stx1618},
  \href {https://ui.adsabs.harvard.edu/abs/2017MNRAS.471.1005M} {471, 1005}

\bibitem[\protect\citeauthoryear{{Malhan} \& {Ibata}}{{Malhan} \&
  {Ibata}}{2018}]{Malhan_SF_2018}
{Malhan} K.,  {Ibata} R.~A.,  2018, \mn@doi [\mnras] {10.1093/mnras/sty912},
  \href {https://ui.adsabs.harvard.edu/abs/2018MNRAS.477.4063M} {477, 4063}

\bibitem[\protect\citeauthoryear{{Malhan} \& {Ibata}}{{Malhan} \&
  {Ibata}}{2019}]{Malhan2019_Pot}
{Malhan} K.,  {Ibata} R.~A.,  2019, \mn@doi [\mnras] {10.1093/mnras/stz1035},
  \href {https://ui.adsabs.harvard.edu/abs/2019MNRAS.486.2995M} {486, 2995}

\bibitem[\protect\citeauthoryear{{Malhan}, {Ibata}  \& {Martin}}{{Malhan}
  et~al.}{2018}]{Malhan_Ghostly_2018}
{Malhan} K.,  {Ibata} R.~A.,   {Martin} N.~F.,  2018, \mn@doi [\mnras]
  {10.1093/mnras/sty2474}, \href
  {https://ui.adsabs.harvard.edu/abs/2018MNRAS.481.3442M} {481, 3442}

\bibitem[\protect\citeauthoryear{{Malhan}, {Ibata}, {Carlberg}, {Bellazzini},
  {Famaey}  \& {Martin}}{{Malhan} et~al.}{2019}]{Malhan2019_Kshir}
{Malhan} K.,  {Ibata} R.~A.,  {Carlberg} R.~G.,  {Bellazzini} M.,  {Famaey} B.,
    {Martin} N.~F.,  2019, \mn@doi [\apjl] {10.3847/2041-8213/ab530e}, \href
  {https://ui.adsabs.harvard.edu/abs/2019ApJ...886L...7M} {886, L7}

\bibitem[\protect\citeauthoryear{{McMillan} \& {Binney}}{{McMillan} \&
  {Binney}}{2010}]{McMillan2010}
{McMillan} P.~J.,  {Binney} J.~J.,  2010, \mn@doi [\mnras]
  {10.1111/j.1365-2966.2009.15932.x}, \href
  {https://ui.adsabs.harvard.edu/abs/2010MNRAS.402..934M} {402, 934}

\bibitem[\protect\citeauthoryear{{Newberg}, {Willett}, {Yanny}  \&
  {Xu}}{{Newberg} et~al.}{2010}]{Newberg2010}
{Newberg} H.~J.,  {Willett} B.~A.,  {Yanny} B.,   {Xu} Y.,  2010, \mn@doi
  [\apj] {10.1088/0004-637X/711/1/32}, \href
  {https://ui.adsabs.harvard.edu/abs/2010ApJ...711...32N} {711, 32}

\bibitem[\protect\citeauthoryear{{Palau} \& {Miralda-Escud{\'e}}}{{Palau} \&
  {Miralda-Escud{\'e}}}{2020}]{Palau2020}
{Palau} C.~G.,  {Miralda-Escud{\'e}} J.,  2020, arXiv e-prints, \href
  {https://ui.adsabs.harvard.edu/abs/2020arXiv201014381P} {p. arXiv:2010.14381}

\bibitem[\protect\citeauthoryear{{Pearson}, {Price-Whelan}  \&
  {Johnston}}{{Pearson} et~al.}{2017}]{Pearson2017}
{Pearson} S.,  {Price-Whelan} A.~M.,   {Johnston} K.~V.,  2017, \mn@doi [Nature
  Astronomy] {10.1038/s41550-017-0220-3}, \href
  {https://ui.adsabs.harvard.edu/abs/2017NatAs...1..633P} {1, 633}

\bibitem[\protect\citeauthoryear{{Petersen} \& {Pe{\~n}arrubia}}{{Petersen} \&
  {Pe{\~n}arrubia}}{2020}]{Petersen2020}
{Petersen} M.~S.,  {Pe{\~n}arrubia} J.,  2020, arXiv e-prints, \href
  {https://ui.adsabs.harvard.edu/abs/2020arXiv201110581P} {p. arXiv:2011.10581}

\bibitem[\protect\citeauthoryear{{Price-Whelan}, {Sesar}, {Johnston}  \&
  {Rix}}{{Price-Whelan} et~al.}{2016}]{PriceWhelan2016}
{Price-Whelan} A.~M.,  {Sesar} B.,  {Johnston} K.~V.,   {Rix} H.-W.,  2016,
  \mn@doi [\apj] {10.3847/0004-637X/824/2/104}, \href
  {https://ui.adsabs.harvard.edu/abs/2016ApJ...824..104P} {824, 104}

\bibitem[\protect\citeauthoryear{{Reid} \& {Brunthaler}}{{Reid} \&
  {Brunthaler}}{2004}]{Reid2004}
{Reid} M.~J.,  {Brunthaler} A.,  2004, \mn@doi [\apj] {10.1086/424960}, \href
  {https://ui.adsabs.harvard.edu/abs/2004ApJ...616..872R} {616, 872}

\bibitem[\protect\citeauthoryear{{Reid} \& {Brunthaler}}{{Reid} \&
  {Brunthaler}}{2020}]{Reid2020}
{Reid} M.~J.,  {Brunthaler} A.,  2020, \mn@doi [\apj]
  {10.3847/1538-4357/ab76cd}, \href
  {https://ui.adsabs.harvard.edu/abs/2020ApJ...892...39R} {892, 39}

\bibitem[\protect\citeauthoryear{{Reid} et~al.,}{{Reid}
  et~al.}{2009}]{Reid2009}
{Reid} M.~J.,  et~al., 2009, \mn@doi [\apj] {10.1088/0004-637X/700/1/137},
  \href {https://ui.adsabs.harvard.edu/abs/2009ApJ...700..137R} {700, 137}

\bibitem[\protect\citeauthoryear{{Reid} et~al.,}{{Reid}
  et~al.}{2014}]{Reid2014}
{Reid} M.~J.,  et~al., 2014, \mn@doi [\apj] {10.1088/0004-637X/783/2/130},
  \href {https://ui.adsabs.harvard.edu/abs/2014ApJ...783..130R} {783, 130}

\bibitem[\protect\citeauthoryear{{Riello, Marco}, {De Angeli, F.}  \& {Evans,
  D. W.}}{{Riello, Marco} et~al.}{2020}]{GaiaEDR3_Riello_2020}
{Riello, Marco} {De Angeli, F.}  {Evans, D. W.} 2020, \mn@doi [A\&A]
  {10.1051/0004-6361/202039587}

\bibitem[\protect\citeauthoryear{{Roederer} \& {Gnedin}}{{Roederer} \&
  {Gnedin}}{2019}]{Roederer2019}
{Roederer} I.~U.,  {Gnedin} O.~Y.,  2019, \mn@doi [\apj]
  {10.3847/1538-4357/ab365c}, \href
  {https://ui.adsabs.harvard.edu/abs/2019ApJ...883...84R} {883, 84}

\bibitem[\protect\citeauthoryear{{Schlafly} \& {Finkbeiner}}{{Schlafly} \&
  {Finkbeiner}}{2011}]{Schlafly2011}
{Schlafly} E.~F.,  {Finkbeiner} D.~P.,  2011, \mn@doi [\apj]
  {10.1088/0004-637X/737/2/103}, \href
  {https://ui.adsabs.harvard.edu/abs/2011ApJ...737..103S} {737, 103}

\bibitem[\protect\citeauthoryear{{Schlegel}, {Finkbeiner}  \&
  {Davis}}{{Schlegel} et~al.}{1998}]{Schlegel1998}
{Schlegel} D.~J.,  {Finkbeiner} D.~P.,   {Davis} M.,  1998, \mn@doi [\apj]
  {10.1086/305772}, \href
  {https://ui.adsabs.harvard.edu/abs/1998ApJ...500..525S} {500, 525}

\bibitem[\protect\citeauthoryear{{Sch{\"o}nrich}, {Binney}  \&
  {Dehnen}}{{Sch{\"o}nrich} et~al.}{2010}]{Schonrich2010}
{Sch{\"o}nrich} R.,  {Binney} J.,   {Dehnen} W.,  2010, \mn@doi [\mnras]
  {10.1111/j.1365-2966.2010.16253.x}, \href
  {https://ui.adsabs.harvard.edu/abs/2010MNRAS.403.1829S} {403, 1829}

\bibitem[\protect\citeauthoryear{{Shipp} et~al.,}{{Shipp}
  et~al.}{2018}]{Shipp2018}
{Shipp} N.,  et~al., 2018, \mn@doi [\apj] {10.3847/1538-4357/aacdab}, \href
  {https://ui.adsabs.harvard.edu/abs/2018ApJ...862..114S} {862, 114}

\bibitem[\protect\citeauthoryear{Sivia}{Sivia}{1996}]{sivia1996data}
Sivia D.,  1996, Data Analysis: A Bayesian Tutorial.
Oxford science publications, Clarendon Press, \url
  {https://books.google.fr/books?id=wR5yljKasLsC}

\bibitem[\protect\citeauthoryear{{Varghese}, {Ibata}  \& {Lewis}}{{Varghese}
  et~al.}{2011}]{Varghese2011}
{Varghese} A.,  {Ibata} R.,   {Lewis} G.~F.,  2011, \mn@doi [\mnras]
  {10.1111/j.1365-2966.2011.19097.x}, \href
  {https://ui.adsabs.harvard.edu/abs/2011MNRAS.417..198V} {417, 198}

\bibitem[\protect\citeauthoryear{{Vasiliev}, {Belokurov}  \&
  {Erkal}}{{Vasiliev} et~al.}{2020}]{Vasiliev2020}
{Vasiliev} E.,  {Belokurov} V.,   {Erkal} D.,  2020, arXiv e-prints, \href
  {https://ui.adsabs.harvard.edu/abs/2020arXiv200910726V} {p. arXiv:2009.10726}

\bibitem[\protect\citeauthoryear{{Yanny} et~al.,}{{Yanny}
  et~al.}{2009}]{Yanny2009}
{Yanny} B.,  et~al., 2009, \mn@doi [\aj] {10.1088/0004-6256/137/5/4377}, \href
  {https://ui.adsabs.harvard.edu/abs/2009AJ....137.4377Y} {137, 4377}

\bibitem[\protect\citeauthoryear{{Zhao}, {Zhao}, {Chu}, {Jing}  \&
  {Deng}}{{Zhao} et~al.}{2012}]{Zhao2012}
{Zhao} G.,  {Zhao} Y.-H.,  {Chu} Y.-Q.,  {Jing} Y.-P.,   {Deng} L.-C.,  2012,
  \mn@doi [Research in Astronomy and Astrophysics]
  {10.1088/1674-4527/12/7/002}, \href
  {https://ui.adsabs.harvard.edu/abs/2012RAA....12..723Z} {12, 723}

\makeatother
\end{thebibliography}
%%%%%%%%%%%%%%%%%%%%%%%%%%%%%%%%%%%%%%%%%%%%%%%%%%

\appendix

\section{Photometric distances of stream stars}\label{appendix:photo_dis}

Here, we describe our method to compute the photometric distances of individual stars in a given stream. Our technique is a two step process: first, we select a SSP template model that suitably represents the stream's observed CMD, and second, we use this template to compute distances of the individual stars. 

\begin{figure}
\begin{center}
\includegraphics[width=\hsize]{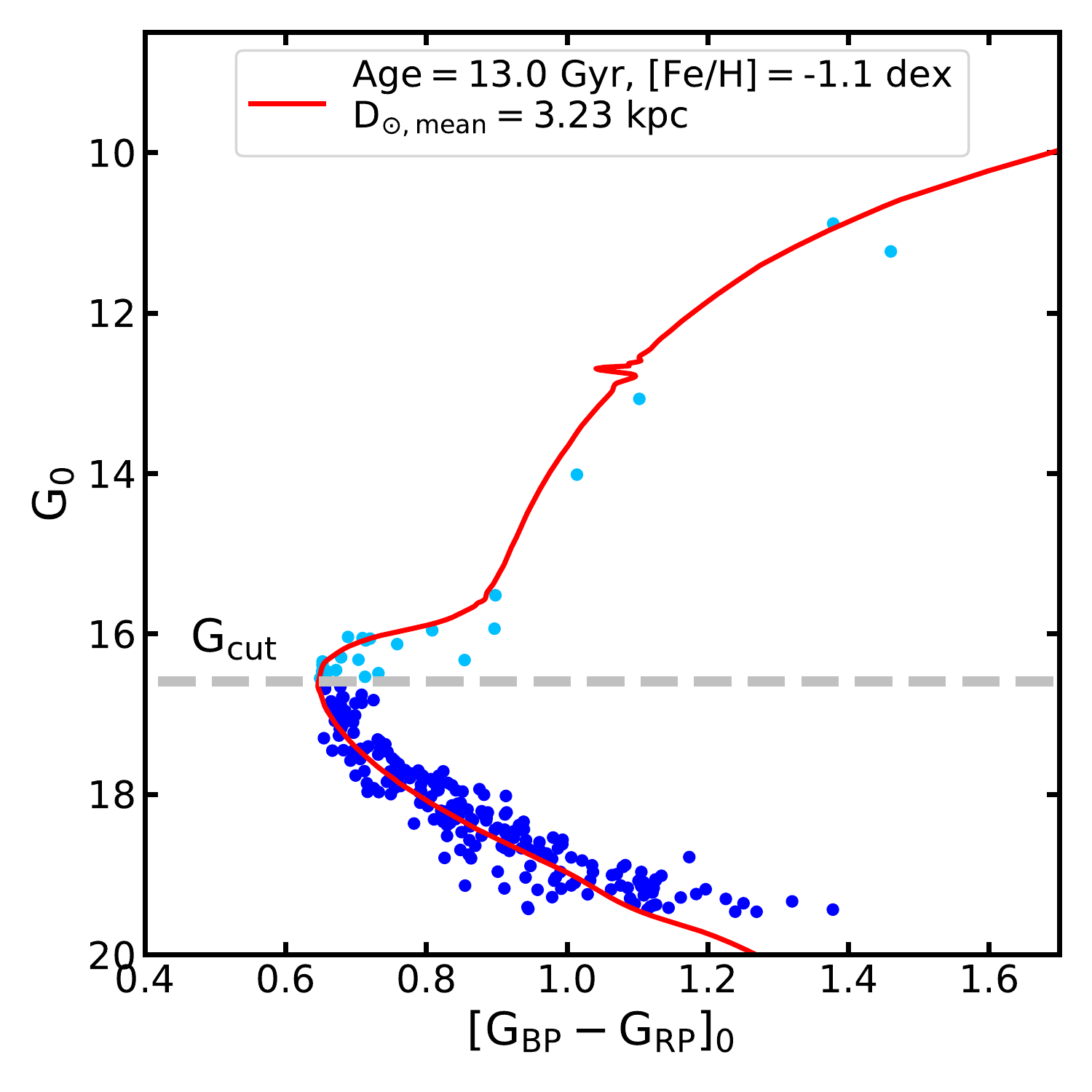}
\end{center}
\vspace{-0.5cm}
\caption{Computing the photometric distances of stream stars. This plot shows color-magnitude distribution of the Hr\'{\i}d stream. The dots show the extinction corrected \Gaia $G$ vs. $G_{\rm BP}-G_{\rm RP}$ photometry of the stars. The stellar population model (red curve) is shifted to the mean of the distance solution found by the \texttt{STREAMFINDER}, which allows for correct calibration. The CMD and the template model are divided along $G-$ magnitude into ``brighter'' and ``fainter'' parts. This cut is made at the MSTO of the model, corresponding to $G_{\rm cut}$. The brighter part of the template is used to compute the distances of the brighter stars, and the fainter part is used to compute the distances of the fainter stars. This pragmatic approach allows us to take into account the distance gradients of the streams.}
\label{fig:Fig_schematic_CMD}
\end{figure}

\begin{enumerate}

\item To find a suitable SSP model for a given stream, we do the following. First, we fix the model [Fe/H] value realistically as described in Section~\ref{sec:data}. After fixing the template model in [Fe/H], we sample over the Age from $10-13\Gyr$ (in bins of $0.5\Gyr$), and compare the corresponding template with the observed CMD ($[G_{\rm BP}-G_{\rm RP}]_0, G_0$). For this comparison, we shift the template along the magnitude direction (in our case, $G$) by $\rm{5log_{10}(D_{\rm \odot, mean})}$, where $D_{\rm \odot, mean}$ corresponds to the mean of the orbital-distance solution found by the \texttt{STREAMFINDER}. This latter point allowed us to correctly calibrate the template model with respect to the CMD (as shown in Fig.~\ref{fig:Fig_schematic_CMD}). While comparing various SSP models, the figure of merit is adopted as the log-likelihood of the data given the model. The log-likelihood function is expressed as
\begin{equation}\label{eq:likelihood_CMD}
\ln \mathcal{L} = \sum_{\rm Data} [-\ln (\sigma_{G} \sigma_{G_{\rm BP}-G_{\rm RP}}) +\ln N -\ln D]\,,
\end{equation}
where
\begin{equation}
\begin{aligned}
N &= \prod_{j=1}^2 (1-e^{-R^2_j/2}) \, , \\
D &= \prod_{j=1}^2 R^2_j \, , \\
R_1^2 &= \Big(\dfrac{ [G_{\rm BP}-G_{\rm RP}]_0^{\rm data} - [G_{\rm BP}-G_{\rm RP}]^{\rm model}}{\sigma_{G_{\rm BP}-G_{\rm RP}}} \Big)^2 \, , \\
R_2^2 &= \Big(\dfrac{ G_0^{\rm data} - G^{\rm shifted\,model}}{\sigma_{G}} \Big)^2 \,.
\end{aligned}
\end{equation}
The quantities $\sigma_{G}$ and $\sigma_{G_{\rm BP}-G_{\rm RP}}$ are the convolution of the intrinsic dispersion of the model together with the observational uncertainty of every star. The reason for avoiding the standard log-likelihood function, and adopting this ``conservative formulation'' of \cite{sivia1996data} was to lower the contribution from the outliers that could be contaminating our data. In sum, this allowed us to find a unique SSP model (of given [Fe/H] and Age) for a given stream.

\item After selecting the unique best fit SSP model, we then compute the distances of stream stars. Since the template is already calibrated at $D_{\rm \odot, mean}$, all we require to do is to compute the relative distance between the stars and the calibrated template. To this end, we do the following. First, we divide both the observed CMD and the calibrated model along $G$-magnitude into ``brighter'' and ``fainter'' parts. This division is made at the main sequence turn-off of the model, corresponding to some value $G_{\rm cut}$. This division is shown in Figure~\ref{fig:Fig_schematic_CMD}. Stars and model points that are brighter than $G_{\rm cut}$ are analysed separately to compute the photometric distances, and those fainter than $G_{\rm cut}$ are analysed separately. To compute the distance relative to $D_{\rm \odot, mean}$, we use the formula
\begin{equation}\label{eq:relative_distance}
\Delta D^i = \dfrac{D_{\rm \odot, mean} (G_0^{\rm i, data} - G^{\rm shifted\,model})}{5{\rm log_{10}}e} \,,
\end{equation}
where $i$ corresponds to a given star, $G_0^{\rm data}$ is the observed $G-$~magnitude of the star, and $G^{\rm shifted\,model}$ is the value from the model. While making this computation, the color measurements and the photometry uncertainties are taken into account. Note that $\Delta D^i$ is positive (negative) if $G_0^{\rm i, data}$ is greater (less) than $G^{\rm shifted\,model}$. Finally, for a given star $i$, Photometric distance=$D_{\rm \odot, mean}+\Delta D^i$. This approach allows to take into account the distance gradients of the streams. 

\end{enumerate}

Although we follow a pragmatic approach, the computed photometric distances of streams were found to be quite similar to the orbital-distance solutions provided by the the \texttt{STREAMFINDER} algorithm.

\section{MCMC chain}\label{appendix:MCMC_chain}

To facilitate future uses of our inferred PDFs, we publish our MCMC chain along with this paper. The file includes $10,000$ randomly selected data points from our full MCMC chain. There are five columns in the file (1) serial number, (2) $\VRsun$, (3) $\Vphisun$, (4) $\Vzsun$ and (5) The value of Log-likelihood as per equation~\ref{eq:likelihood}. Here, the velocities are in the units of $\kms$. We encourage the reader to use this MCMC chain as an input to future analyses instead of the simplified 1-dimensional constraints provided in the abstract. This will ensure that the correlations between the different model parameters are fully taken into account (see Figure~\ref{fig:Corner_Plot_2}).

% Don't change these lines
\bsp	% typesetting comment
\label{lastpage}
\end{document}